# Optical Spectra of Candidate International Celestial Reference Frame (ICRF) Flat-Spectrum Radio Sources. III

O. Titov

Geoscience Australia, PO Box 378, Canberra, ACT 2601, Australia

oleg.titov@ga.gov.au

T. Pursimo

Nordic Optical Telescope, Nordic Optical Telescope Apartado 474E-38700 Santa Cruz de La Palma, Santa Cruz de Tenerife, Spain

Helen M. Johnston

Sydney Institute for Astronomy, School of Physics, University of Sydney, NSW 2006, Australia

Laura M. Stanford

Geoscience Australia, PO Box 378, Canberra, ACT 2601, Australia

Richard W. Hunstead

Sydney Institute for Astronomy, School of Physics, University of Sydney, NSW 2006, Australia

David L. Jauncey

CSIRO Astronomy and Space Science, ATNF & Mount Stromlo Observatory, Cotter Road, Weston, ACT 2611, Australia

and

Katrina A. Zenere

School of Physics, University of Sydney, NSW 2006, Australia

## ABSTRACT

In extending our spectroscopic program, which targets sources drawn from the International Celestial Reference Frame (ICRF) Catalog, we have obtained spectra for ∼160 compact flat-spectrum radio sources and determined redshifts for 112 quasars and radio galaxies. A further 14 sources with featureless spectra have been classified as BL Lac objects. Spectra were obtained at three telescopes: the 3.58 m ESO New Technology Telescope (NTT), and the two 8.2 m Gemini telescopes in Hawaii and Chile. While most of the sources are powerful quasars, a significant fraction of radio galaxies is also included from the list of non-defining ICRF radio sources.

*Subject headings:* reference systems – galaxies: active – quasars: emission lines – BL Lacertae objects: general – radio continuum: general

## 1. Introduction

The accurate alignment of astronomical catalogues produced at different wavelengths is critical for confident identification of celestial objects. Milliarcsecond positions of compact extragalactic radio sources have been used since 1998 to define the International Celestial Reference Frame (ICRF; Ma, Arias & Eubanks 1998). The current realization of the International Celestial Reference System (ICRS; Arias et al. 1995) is the catalogue known as ICRF2 (Second Realization of the International Celestial Reference Frame). It is based on the progressive refinement since 1984 of the positions of a set of reference radio sources measured with Very Long Baseline Interferometry (VLBI) techniques (Ma et al. 2009, Fey et al. 2015). Some technical details are provided in our previous papers (Titov et al. 2011, 2013). The ICRF2 catalogue is based on the assumption that the reference radio sources have no significant proper motion, so their astrometric positions are affected only by the relativistic jets. However, at microarcsec levels Bastian (1995), Gwinn et al. (1997), Sovers, Fanselow & Jacobs (1998), and Kovalevsky (2003) have drawn attention to the effect of aberration in proper motions, while a dipole systematic of magnitude $\sim 6$ $\mu$as/year, detected by Titov, Lambert & Gontier (2011), Xu, Wang & Zhao (2012) and Titov & Lambert (2013), may degrade positions in the same way that proper motions of stars degraded the old optical catalogs.

The astrometric mission *Gaia* (Perryman et al. 2001; Mignard 2012), was launched on 2013 December 19 by the European Space Agency, and the first results have recently been published (Lindegren et al. 2016). One of the mission goals has been to measure high precision positions and proper motions of $\sim$500,000 quasars. A new astrometric reference frame based on the optical counterparts of the ICRF radio sources is being built, assuming that quasar parallaxes are zero (Michalik & Lindegren 2016). These new data will also test for the presence and magnitude of the dipole proper motion reported by Titov, Lambert

& Gontier (2011), Xu, Wang & Zhao (2012) and Titov & Lambert (2013), as well as any higher-order systematics. The alignment of radio and optical reference frames demands a reliable optical identification of the radio sources and, if possible, a redshift to confirm the identification. However, only approximately half of the ICRF radio sources have a measured redshift (Titov & Malkin 2009).

The reference radio sources are either optically bright quasars or extended radio galaxies. Our ongoing observing strategy first identifies the optical counterpart for each source, using either the SuperCOSMOS Sky Survey (Hambly et al. 2001) or the Sloan Digital Sky Survey (DR9 through 13) (SDSS; York et al. 2000), and we then obtain a spectrum to secure a redshift. We routinely obtain a short-exposure wideband image of the field prior to starting the spectroscopic exposure, firstly to confirm the identification and secondly to guard against possible confusion by foreground stars, especially in crowded fields (e.g., see Titov et al. 2013, Fig. 3); good seeing is a critical issue in the latter case.

The current paper, Part III of our ongoing project, presents redshifts for 112 targets, mostly in the southern hemisphere. The majority appear point-like in the optical with broad permitted lines or BL Lac objects with featureless spectra; only two clear galaxies with narrow emission lines are presented, although classification of images near the limiting magnitudes of the optical surveys is difficult. We describe the observations and data reduction procedures in Section 2. Our spectra, along with detailed comments on individual sources, are presented in Section 3. A discussion of targets showing interesting emission or absorption properties is given in Section 4, followed by a summary and conclusion in Section 5. Where relevant, a $\Lambda$CDM cosmology has been assumed, with $H_0 = 70 \text{ km s}^{-1} \text{ Mpc}^{-1}$, $\Omega_M = 0.29$ and $\Omega_\Lambda = 0.71$.

## 2. Observations

Spectroscopic observations were carried out at three optical facilities.

*ESO NTT*. We had a 5-night observing run in Visitor Mode at the European Southern Observatory (ESO) 3.58 m New Technology Telescope (NTT) in 2013 December (Proposal 092.A-0021 (A)) using the ESO Faint Object Spectrograph and Camera (EFOSC) system with grism #13 covering the wavelength range 3685–9315 Å. The seeing during observations was typically $0\rlap{.}''5 - 2\rlap{.}''0$, with a spectral resolution 21 Å FWHM. Exposure times varied from 5–30 minutes depending on the magnitude of each target and current sky conditions. Wavelength calibration made use of HeNeAr comparison spectra, resulting in an rms accuracy of 0.5 Å.

*Gemini*: A large number of targets were observed in Service Mode at the Gemini North and Gemini South 8.2 m telescopes through the Poor Weather Program (Proposals GN-2012B-Q-127, GS-2013A-Q-99, GS-2014A-Q-93) using the Gemini Multi-Object Spectrograph (GMOS) system with grating R400 at each site. This grating covers 4500 Å centered at 5200 Å. The wavelength resolution was ∼15 Å FWHM, and an exposure time of 20 minutes was used for all targets. Wavelength calibration used the spectra of a CuAr comparison lamp, giving an rms accuracy of 0.3 Å.

Data reduction was performed with the IRAF software suite[1] using standard procedures for spectral analysis. Bias and flat-field correction was applied to each frame, and cosmic rays were removed using the IRAF task SZAP. Where more than one exposure was obtained, the separate exposures were combined. After spectrum extraction, sky subtraction and wavelength calibration, the final one-dimensional spectra were flux-calibrated with a spectrophotometric standard observed with the same instrumental setup. Because the observing conditions were often non-photometric, especially for Gemini observations made through the Poor Weather Program, the flux calibration should be taken as approximate.

---

[1]IRAF is distributed by the National Optical Astronomical Observatory, which is operated by the Association of Universities for Research in Astronomy, Inc., under contract to the National Science Foundation.

## 3. Results

We present our spectroscopic results in the same format as our previous paper (Titov et al. 2013). The redshifts of 112 IVS objects are listed in Table 1, along with their J2000 ICRF2 positions (Ma et al. 2009, Fey et al. 2015), the telescope used for each spectrum, the identified emission lines (rest and observed wavelengths), the mean redshift and error, and brief notes on individual sources. Additional notes on individual sources, marked with an asterisk in the final column, are given in Section 3.1. The quoted errors in the mean redshift $\overline{z}$ were calculated as discussed in Titov et al. (2013), i.e., from the quadratic combination of the scatter in the redshift estimates and the fractional error in the wavelength calibration. Single-line redshifts, usually Mg II, are assigned an uncertainty of $\Delta z = 0.001$ if the signal-to-noise (S/N) is high, increasing to 0.003 for low S/N, or a broad or asymmetric line.

Figure 1 displays the individual spectra, along with the line identifications. Blue dashed lines indicate lines that were used for redshift calculation, while red dot-dashed line indicates lines that were present, generally at a low S/N, but not used in determining the mean redshift. Telluric lines near 6850 Å and 7600 Å are marked on the spectra.

Fourteen objects with good S/N ratio and featureless spectra were classed as probable BL Lac objects. They are listed in Table 2 with their ICRF2 positions (Ma et al. 2009, Fey et al. 2015), and their spectra are shown in Figure 2. Comments on two of them are included in Section 3.1. There is also a significant fraction of quasars in Fig. 1 with weak emission lines on a strong continuum (e.g. IVS B0530−484 and B0619−468), which we assume are incipient BL Lacs, perhaps not surprising in a VLBI-selected sample.

An additional 23 targets had low S/N spectra that did not permit a confident spectral classification. These are listed in Table 3 with their ICRF2 positions. Attempts to identify and obtain redshifts for 11 IVS targets at low Galactic latitude were not successful.

### 3.1. Notes on individual targets

1. IVS B0023−354—intervening absorption system at $z_{abs} = 1.3167 \pm 0.0010$ with lines of Fe II $\lambda\lambda2586$, 2600, Mg II $\lambda\lambda2796$, 2803 and Mg I $\lambda2852$. Although the Mg II doublet is not resolved, the presence of Mg I and Fe II confirm the system.

2. IVS B0048+447—two associated absorption systems are detected within the Mg II emission line, one at $z_{abs} = 1.2568 \pm 0.0001$ and the other, including Mg I $\lambda2852$, at $z_{abs} = 1.2715 \pm 0.0001$. Their velocities relative to the Mg II emission line are $-370$ and $+2300\,km\,s^{-1}$, respectively.

3. IVS B0229+262—a possible intervening C IV $\lambda\lambda1548$, 1550 doublet is detected at $z_{abs} = 2.5261 \pm 0.0001$.

4. IVS B0307−085—very strong associated absorption in C IV $\lambda\lambda1548$, 1550, Si IV $\lambda\lambda1393$, 1402, N V $\lambda\lambda1238$, 1242 and Ly$\alpha$ $\lambda1215$ at a redshift of $z_{abs} = 3.2899 \pm 0.0005$, consistent with zero velocity shift relative to emission.

5. IVS B0521−403—probable intervening absorption system at $z_{abs} = 0.6974 \pm 0.0001$ based on Mg II $\lambda\lambda2796$, 2803 (blend) and Mg I $\lambda2852$ lines. Spectral resolution is too low to separate the Mg II doublet, but the wavelength match is good. There are two further absorption lines at 5023Å and 5997Å, which could not be identified but may be additional blended Mg II doublets based on their line widths.

6. IVS B0550−453—intervening absorption system at $z_{abs} = 0.6859 \pm 0.0001$ based on lines of Fe II $\lambda\lambda2344$, 2382, 2586+2600 (blend) and Mg II $\lambda\lambda2796$, 2803 (blend).

7. IVS B0722−068—broad emission from Fe II multiplets both blueward and redward of H$\beta$ (see Fig. 5).

8. IVS B0727−365—permitted lines Mg II, H$\gamma$ and H$\beta$ are displaced ∼1200 km s$^{-1}$ to higher redshift relative to the forbidden lines; the overall mean redshift is given in Table 1.

9. IVS B0838−573—absorption line at 6237 Å is a possible intervening Mg II doublet at $z_{abs} = 1.2304$.

10. IVS B0858−313—broad emission from Fe II multiplets both blueward and redward of H$\beta$; [O III] $\lambda5007$ may be partly blended with Fe II $\lambda5018$ (see Fig. 5).

11. IVS B0917−354—broad emission from Fe II multiplets both blueward and redward of H$\beta$; [O III] $\lambda5007$ may be partly blended with Fe II $\lambda5018$ (see Fig. 5).

12. IVS B1006−093—probable blend of associated Mg II $\lambda\lambda2796$, 2803 absorption at $z_{abs} = 1.1884 \pm 0.0001$.

13. IVS B1219−457—prominent broad absorption line in the blue wing of Mg II emission at $z_{abs} = 1.034$, corresponding to an outflow velocity of ∼2600 km s$^{-1}$.

14. IVS B1228−352—absorption line at 4151 Å is a possible intervening Mg II doublet at $z_{abs} = 0.4845$.

15. IVS B1229−123—self-absorption in C IV $\lambda\lambda1548$, 1550 is matched by absorption in Ly$\alpha$ $\lambda1215$ at $z_{abs} = 2.1054 \pm 0.0003$.

16. IVS B1418−481—intervening absorption system at $z_{\mathrm{abs}} = 1.0393 \pm 0.0001$ is based on Mg II $\lambda\lambda 2796$, 2803 and Mg I $\lambda 2853$.

17. IVS B1558−072—bright quasar with an intervening absorption system at $z_{\mathrm{abs}} = 1.0787 \pm 0.0001$ based on lines of Fe II $\lambda\lambda 2344$, 2382, 2586, 2600 and Mg II $\lambda\lambda 2796$, 2803.

18. IVS B1711−670 (Fig. 2)—observed in poor seeing and we assume that the close star $2\overset{''}{.}5$ to the south-west is responsible for contamination by weak H$\alpha$ and H$\beta$ absorption at rest. Spectrum otherwise is featureless and the target is tentatively classified as a BL Lac object.

19. IVS B1712+530—bright quasar with an intervening absorption system at $z_{\mathrm{abs}} = 1.5145 \pm 0.0010$ based on lines of Fe II $\lambda\lambda 2586$, 2600 and Mg II $\lambda\lambda 2796$, 2803.

20. IVS B1810−745—self-absorption near the centre of C IV emission at $z_{\mathrm{abs}} = 2.246$.

21. IVS B1813−497—spectrum was obtained in poor seeing and is partly contaminated by light from a close star $\sim 0.8$ arcsec west of the quasar; see Fig. 3.

22. IVS B1905−252—intervening absorption system at $z_{\mathrm{abs}} = 1.3165 \pm 0.0001$ based on Fe II $\lambda 2600$ and Mg II $\lambda\lambda 2796$, 2803.

23. IVS B1957−423—absorption system, probably associated, at $z = 0.9674 \pm 0.0001$ in the extreme blue wing of Mg II emission, based on Mg II $\lambda\lambda 2796$, 2803 and Fe II $\lambda 2600$; the blueshift is $\sim 3200$ km s$^{-1}$.

24. IVS B2029−215—an FeLoBAL quasar, with very strong broad absorption and emission, predominantly in Fe II; see Section 4.3 and Fig. 6.

25. IVS B2036−034—intervening absorption system at $z_{\mathrm{abs}} = 1.2097 \pm 0.0010$ based on lines of Fe II $\lambda\lambda 2344$, 2382, 2586, 2600 and Mg II $\lambda\lambda 2796$, 2803.

26. IVS B2044−188—intervening absorption system at $z_{abs} = 1.4905 \pm 0.0010$ based on lines of Fe II $\lambda\lambda2344$, 2382, 2586, 2600, Mg II $\lambda\lambda2796$, 2803 and Mg I $\lambda2852$.

27. IVS B2107−553—the two absorption lines, at 4747.2Å and 4801.1Å can not be identified with any known doublet or pair of lines.

28. IVS B2131−690—very broad emission lines of C III] and Mg II.

29. IVS B2132−638—prominent associated Mg II $\lambda\lambda2796$, 2803 absorption close to the peak of the emission line at $z = 1.5349 \pm 0.0001$ with a blueshift of $\sim250$ km s$^{-1}$.

30. IVS B2155−475—two absorption line systems: an intervening Mg II $\lambda\lambda2796$, 2803 system at $z = 0.7292 \pm 0.0044$, and an associated system at $z = 1.7423 \pm 0.0010$ based on Si II $\lambda1526$, C IV $\lambda\lambda1548$, 1550 (blend), C I $\lambda1560$ and Fe II $\lambda1608$.

31. IVS B2257+313 (Fig. 2)—featureless spectrum with a clear absorption system at $z_{abs} = 0.9317 \pm 0.0001$ defined by lines of Fe II $\lambda\lambda2586$, 2600 and Mg II $\lambda\lambda2796$, 2803, thereby setting a lower limit on the redshift of this BL Lac object.

32. IVS B2300+386—two intervening absorption systems at $z_{abs} = 1.0138 \pm 0.0001$ and $z_{abs} = 1.6005 \pm 0.0001$ based solely on the Mg II $\lambda\lambda2796$, 2803 doublet.

33. IVS B2313−182—low-luminosity radio galaxy with narrow [O II] $\lambda3727$ and [O III] $\lambda\lambda4959$, 5007 emission on a predominantly stellar continuum with clear Ca II absorption.

### 3.2. Separation of close objects

Galactic stars are occasionally found close on the sky to the VLBI radio position, either contaminating the spectrum or so closely aligned with the radio source that the field is completely obscured. Of the three examples shown in Fig. 3, redshifts were able to be recovered for IVS B0905−202 and B1813−497, although the latter spectrum, taken in poor seeing, is clearly contaminated by stellar absorption lines at rest. In the case of IVS B0758−737, which is relatively bright on VLBI baselines, there is also the opportunity to monitor changes with time in the VLBI position as proper motion of the foreground star changes the gravitational deflection of the quasar light.

## 4. Discussion

The longer integration times in the present dataset have brought to light some spectral features, both in emission and absorption, that were either not detectable in the earlier datasets or extremely rare (Titov et al. 2011, 2013). These are discussed briefly in the following sections.

### 4.1. Mg II absorption systems

The generally good S/N has allowed the detection of a number of absorption systems, both those associated with the target quasar and those intervening along the line of sight. The resolution of the NTT spectra is not sufficient to separate the Mg II doublet, but in most cases the system is confirmed by the presence of other lines, notably Mg I and/or Fe II. Absorption line redshifts were determined using SPLOT or SPECFIT in IRAF. Gaussian profiles were fitted to the lines, assuming a single redshift and width for all lines. Redshifts span the range $z_{abs} = 0.7$–$1.5$.

A montage of the Mg II systems shifted to the rest frame is shown in Fig. 4. The NTT spectral resolution is too low to separate the Mg II doublet and Fig. 4 only includes those NTT systems with confirming lines of Mg I and/or Fe II. Mg II doublet ratios in the Gemini spectra are all close to 1, indicating that the doublet is at, or close to, saturation with equivalent widths $W_0^{\lambda 2796} > 0.5$Å.

The principal interest in IVS quasars showing strong Mg II absorption, both associated and intervening, is their value as potential targets in searches for redshifted neutral hydrogen, taking advantage of the similar excitation potentials (and therefore locations) of H I and Mg II. While our sample is small in comparison with those from the Sloan Digital Sky Survey (eg. Nestor et al. 2005), the majority of the Sloan sample is radio quiet. Moreover, ICRF source selection ensures that a substantial fraction of the radio flux density is contained within a few milliarcsec, thereby increasing the likelihood of an H I detection. Source selection played an important role in the first successful detection of redshifted H I with the Australian SKA Pathfinder (ASKAP; Johnston et al. 2007) where the target was the peaked spectrum VLBI source IVS B1740−517 (Allison et al. 2015) at $z = 0.44$; unfortunately, the optical spectrum, obtained at Gemini South through Director's Discretionary Time, did not cover the Mg II line.

## 4.2. Fe II emission near H$\beta$

Three of the low-$z$ quasars show prominent Fe II multiplet emission near H$\beta$: IVS B0722−068, B0858−313 and B0917−354. A montage of these three spectra, reduced to the emission rest frame, is shown in Fig. 5. Fe II emission in this spectral region was first noted and analysed by Wampler & Oke (1967) in the spectrum of 3C 273. They noted how the $\lambda 4924$ line could distort the red wing of the H$\beta$ profile and the $\lambda 5018$ line could blend with [O III] $\lambda 5007$.

### 4.3. The FeLoBAL quasar IVS B2029−215

An extreme example of broad Fe II emission and absorption is shown in the spectrum of the quasar, IVS B2029−215, which is dominated by broad Fe II emission and absorption lines. This object belongs to a rare class of low-ionization broad absorption line quasars known as FeLoBALs, so-named because the spectrum is dominated by excited states of Fe II. The spectrum, shifted to the emission rest frame, is shown in Fig. 6.

The spectrum of IVS B2029−215 is very similar to that of the $z = 0.692$ quasar FIRST J121442.3+280329 (de Kool et al. 2002; Branch et al. 2002) from the FIRST Bright Quasar Survey (White et al. 2000). By analogy with the spectrum of FIRST J121442.3+280329 (de Kool et al. 2002) we show a tentative continuum in Fig. 6 on the assumption that broad emission and blue-shifted broad absorption are superimposed on a power-law continuum from the accretion disk. The IVS B2029−215 continuum, however, is substantially redder than that of FIRST J121442.3+280329, suggesting the presence of dust in the outflowing gas.

### 4.4. Continuum absorption below Lyα

It is well known that the continuum level in high redshift quasars is depressed below Lyα relative to an extrapolation of the level above Lyα. The fractional depression, $D_A$, of the continuum between Lyα and Lyβ increases with redshift due mainly to the increasing density of absorption lines of the Lyα forest, with additional contributions from heavy-element lines and, at the highest redshifts, distributed neutral hydrogen (e.g. Jenkins & Ostriker 1991.)

In the present sample there are 8 quasars with $z > 2.6$ and good spectral coverage down to Lyβ. Of these, five show the expected continuum depression below Lyα but three,

IVS B0407−129, B1020−045 and B1044−512, have $D_A \approx 0$. The latter three spectra, all at $z \sim 2.6$, were obtained at the NTT in visitor mode and the spectral response has been well calibrated with standard star observations. The implication is that these sightlines have a lower density of Ly$\alpha$ forest lines, and/or a more highly ionized intergalactic medium. To follow up these possibilities, we are planning to observe the region well below Ly$\alpha$ in these quasars at higher spectral resolution and signal-to-noise.

## 5. Summary and conclusion

We present redshifts and spectra of a further 112 emission-line objects identified with VLBI radio sources drawn from the candidate International Celestial Reference Catalog. As in our previous papers (Titov et al. 2011, 2013) the chosen targets were mostly in the south so as to improve the uniformity across the two hemispheres. Redshifts were generally based on two or more emission lines, but single-line redshifts, all Mg II, were considered reliable on the basis of other spectral information. There were 14 objects with high S/N but featureless spectra that we classified as probable BL Lac objects, while a further 22 had S/N too low for reliable classification.

The majority of spectra show the broad permitted lines and power-law continua characteristic of quasars, but the sample does include two low-redshift galaxies with only narrow forbidden lines and stellar continua. Morphological differentiation is difficult at redshifts $z > 0.5$.

Amongst the sample of 112 spectra, we identify 17 low-redshift Mg II absorbers, both associated and intervening. Because these absorbers are seen against very compact radio sources, they are potential targets for H I absorption searches with next generation radio telescopes.

This paper is based on data collected at three telescopes:

1. ESO NTT, under the European Organisation for Astronomical Research in the Southern Hemisphere, Chile under program 092.A-0021 (A)

2. Two Gemini Observatories, which are operated by the Association of Universities for Research in Astronomy, Inc., under a cooperative agreement with the National Science Foundation on behalf of the Gemini partnership: the National Science Foundation (United States), the Science and Technology Facilities Council (United Kingdom), the National Research Council (Canada), CONICYT (Chile), the Australian Research Council (Australia), Ministério da Ciência, Tecnologia, Inovações e Comunicações (Brazil) and Ministerio de Ciencia, Tecnología e Innovación Productiva (Argentina) under programs GS-2013A-Q-99 and GS-2014A-Q-93 (Gemini South), and GN-2012B-Q-127 (Gemini North).

Funding for SDSS-III has been provided by the Alfred P. Sloan Foundation, the Participating Institutions, the National Science Foundation and the U.S. Department of Energy Office of Science. The SDSS-III web site is http://www.sdss3.org/.

SDSS-III is managed by the Astrophysical Research Consortium for the Participating Institutions of the SDSS-III Collaboration including the University of Arizona, the Brazilian Participation Group, Brookhaven National Laboratory, University of Cambridge, Carnegie Mellon University, University of Florida, the French Participation Group, the German Participation Group, Harvard University, the Instituto de Astrofisica de Canarias, the Michigan State/Notre Dame/JINA Participation Group, Johns Hopkins University, Lawrence Berkeley National Laboratory, Max Planck Institute for Astrophysics, Max Planck Institute for Extraterrestrial Physics, New Mexico State University, New York University, Ohio State University, Pennsylvania State University, University of Portsmouth,

Table 1.   Observed emission lines

| Source | RA (J2000)[a] | Dec (J2000)[a] | Telescope[b] | Line | Rest $\lambda$ | Obs $\lambda$[c] | $z$ | Note[d] |
|---|---|---|---|---|---|---|---|---|
| IVS B0001−440 | 00 04 07.2575 | −43 45 10.146 | GS | Mg II | 2799.9 | 5582.9 | $0.994 \pm 0.001$ | S * |
| IVS B0009+081 | 00 11 35.2696 | +08 23 55.586 | GN | C III] | 1908.7 | 4587.4 | $1.4032 \pm 0.0003$ | |
| | | | | Mg II | 2799.9 | 6728.1 | | |
| IVS B0010−873 | 00 11 51.4792 | −87 06 25.453 | GS | Ly$\alpha$ | 1215.7 | 5218.9 | $3.2928 \pm 0.0026$ | |
| | | | | N V | 1240.1 | 5329.9 | | |
| | | | | Si II | 1304.4 | 5601.8 | | |
| | | | | C II | 1334.5 | [5737.4] | | |
| | | | | Si IV | 1398.3 | [5985.2] | | |
| | | | | C IV | 1549.5 | 6640.6 | | |
| IVS B0016−223 | 00 19 22.8280 | −22 05 27.938 | GS | Mg II | 2799.9 | 4637.1 | $0.6559 \pm 0.0003$ | |
| | | | | [Ne V] | 3345.4 | 5536.8 | | |
| | | | | [Ne V] | 3425.5 | 5674.5 | | |
| | | | | [O II] | 3726.8 | 6172.2 | | |
| | | | | [Ne III] | 3868.7 | 6405.2 | | |
| | | | | [Ne III] | 3967.4 | [6571.1] | | |
| | | | | H$\delta$ | 4101.7 | [6789.1] | | |

| Source | RA (J2000)[a] | Dec (J2000)[a] | Telescope[b] | Line | Rest $\lambda$ | Obs $\lambda$[c] | $z$ | Note[d] |
|---|---|---|---|---|---|---|---|---|
| | | | | H$\gamma$ | 4340.5 | [7182.6] | | |
| | | | | [O III] | 4363.2 | [7219.4] | | |
| IVS B0023−354 | 00 26 16.3871 | −35 12 48.789 | NTT | C IV | 1549.5 | 4625.7 | $1.996 \pm 0.005$ | * |
| | | | | C III] | 1908.7 | 5730.0 | | |
| | | | | Mg II | 2799.9 | 8397.9 | | |
| IVS B0048+447 | 00 51 36.4736 | +44 59 35.958 | GN | Mg II | 2799.9 | 6311.7 | $1.254 \pm 0.001$ | S * |
| IVS B0105+129 | 01 07 45.9618 | +13 12 05.191 | GN | Mg II | 2799.9 | 4929.9 | $0.7600 \pm 0.0002$ | |
| | | | | [Ne V] | 3345.4 | 5885.5 | | |
| | | | | [Ne V] | 3425.5 | 6028.7 | | |
| | | | | [O II] | 3726.8 | 6558.7 | | |
| | | | | [Ne III] | 3868.7 | 6808.6 | | |
| | | | | [Ne III] | 3967.4 | 6982.7 | | |
| | | | | H$\delta$ | 4101.7 | [7217.5] | | |
| IVS B0117−171 | 01 19 43.6459 | −16 54 08.972 | GS | Mg II | 2799.9 | 4881.3 | $0.7440 \pm 0.0005$ | |
| | | | | [Ne V] | 3425.5 | [5980.1] | | |
| | | | | [O II] | 3726.8 | 6503.4 | | |

| Source | RA (J2000)[a] | Dec (J2000)[a] | Telescope[b] | Line | Rest $\lambda$ | Obs $\lambda$[c] | $z$ | Note[d] |
|---|---|---|---|---|---|---|---|---|
| | | | | [Ne III] | 3868.7 | 6745.1 | | |
| IVS B0134+215 | 01 37 15.6249 | +21 45 44.270 | GN | Mg II | 2799.9 | 5842.7 | $1.0864 \pm 0.0004$ | |
| | | | | [Ne V] | 3425.5 | 7145.6 | | |
| IVS B0137−273 | 01 39 55.3988 | −27 05 29.107 | NTT | Ly$\alpha$ | 1215.7 | [5097.2] | $3.147 \pm 0.005$ | |
| | | | | Si IV | 1398.3 | 5811.4 | | |
| | | | | C IV | 1549.5 | 6412.1 | | |
| | | | | C III] | 1908.7 | 7912.0 | | |
| IVS B0158+096 | 02 01 15.4974 | +09 54 54.392 | GN | Mg II | 2799.9 | 6324.8 | $1.259 \pm 0.001$ | S |
| IVS B0159−668 | 02 01 07.7442 | −66 38 12.665 | NTT | C III] | 1908.7 | 4339.2 | $1.279 \pm 0.006$ | |
| | | | | Mg II | 2799.9 | 6394.2 | | |
| IVS B0200+30A | 02 03 44.2706 | +30 42 37.836 | GN | Mg II | 2799.9 | 4928.3 | $0.7598 \pm 0.0002$ | |
| | | | | [Ne V] | 3425.5 | 6026.3 | | |
| | | | | [O II] | 3726.8 | 6559.3 | | |
| | | | | [Ne III] | 3868.7 | 6806.6 | | |
| | | | | [Ne III] | 3967.4 | 6982.2 | | |
| IVS B0224−189 | 02 26 47.6283 | −18 43 39.238 | GS | C IV | 1549.5 | 4129.6 | $1.679 \pm 0.013$ | |

| Source | RA (J2000)[a] | Dec (J2000)[a] | Telescope[b] | Line | Rest $\lambda$ | Obs $\lambda$[c] | $z$ | Note[d] |
|---|---|---|---|---|---|---|---|---|
| | | | | C III] | 1908.7 | 5138.1 | | |
| IVS B0228−624 | 02 30 11.5749 | −62 14 35.714 | NTT | Si IV | 1398.3 | [3799.1] | $1.6932 \pm 0.0026$ | |
| | | | | C IV | 1549.5 | 4176.5 | | |
| | | | | C III] | 1908.7 | 5130.6 | | |
| | | | | Mg II | 2799.9 | 7548.7 | | |
| IVS B0229+262 | 02 32 27.6232 | +26 28 38.590 | GN | Ly$\alpha$ | 1215.7 | 4584.3 | $2.7720 \pm 0.0005$ | |
| | | | | N V | 1240.1 | [4683.2] | | |
| | | | | Si IV | 1398.3 | [5273.4] | | |
| | | | | C IV | 1549.5 | 5845.7 | | |
| | | | | He II | 1640.4 | 6187.1 | | |
| | | | | C III] | 1908.7 | 7201.1 | | |
| IVS B0230−288 | 02 32 36.7412 | −28 39 26.306 | NTT | Mg II | 2799.9 | 5607.4 | $1.003 \pm 0.002$ | S |
| | | | | H $\delta$ | 4101.7 | [8230.9] | | |
| IVS B0234+469 | 02 38 16.5032 | +47 12 18.551 | GN | Mg II | 2799.9 | 4956.2 | $0.7695 \pm 0.0007$ | |
| | | | | [O II] | 3726.8 | 6592.0 | | |
| | | | | H$\delta$ | 4101.7 | [7262.3] | | |

| Source | RA (J2000)[a] | Dec (J2000)[a] | Telescope[b] | Line | Rest $\lambda$ | Obs $\lambda$[c] | $z$ | Note[d] |
|---|---|---|---|---|---|---|---|---|
| IVS B0244+327 | 02 47 41.2322 | +32 54 31.836 | GN | C III] | 1908.7 | 5185.0 | $1.7165 \pm 0.0003$ | |
| | | | | Mg II | 2799.9 | [7589.0] | | |
| IVS B0307−085 | 03 10 02.0490 | −08 23 39.297 | GS | Ly$\alpha$ | 1215.7 | 5224.5 | $3.289 \pm 0.005$ | * |
| | | | | N V | 1240.1 | [5324.5] | | |
| | | | | Si IV | 1398.3 | 5998.1 | | |
| | | | | C IV | 1549.5 | 6630.5 | | |
| IVS B0307−362 | 03 09 02.5107 | −36 04 03.746 | GS | C III] | 1908.7 | 4206.1 | $1.2051 \pm 0.0015$ | |
| | | | | Mg II | 2799.9 | 6178.3 | | |
| IVS B0322−555 | 03 24 05.5484 | −55 20 54.140 | GS | Ly$\alpha$ | 1215.7 | 3861.9 | $2.1780 \pm 0.0007$ | |
| | | | | Si IV | 1398.3 | 4442.6 | | |
| | | | | C IV | 1549.5 | 4923.4 | | |
| | | | | He II | 1640.4 | 5212.5 | | |
| | | | | Si III] | 1890.0 | 6012.5 | | |
| | | | | C III] | 1908.7 | 6065.6 | | |
| IVS B0324−379 | 03 26 03.0004 | −37 46 09.181 | NTT | Mg II | 2799.9 | 5247.1 | $0.8749 \pm 0.0014$ | |
| | | | | [Ne V] | 3425.5 | 6418.9 | | |

| Source | RA (J2000)[a] | Dec (J2000)[a] | Telescope[b] | Line | Rest $\lambda$ | Obs $\lambda$[c] | $z$ | Note[d] |
|---|---|---|---|---|---|---|---|---|
| | | | | [O II] | 3726.8 | 6979.3 | | |
| | | | | H$\gamma$ | 4340.5 | 8156.0 | | |
| IVS B0325−182 | 03 27 43.3416 | −18 03 42.027 | NTT | Mg II | 2799.9 | 4841.2 | $0.7303 \pm 0.0004$ | |
| | | | | H$\delta$ | 4101.7 | 7099.0 | | |
| | | | | H$\gamma$ | 4340.5 | 7513.9 | | |
| | | | | H$\beta$ | 4861.3 | 8413.7 | | |
| | | | | [O III] | 5006.8 | 8659.6 | | |
| IVS B0407−129 | 04 10 13.9529 | −12 51 11.209 | NTT | Ly$\beta$ | 1025.7 | [3768.0] | $2.649 \pm 0.004$ | |
| | | | | Ly$\alpha$ | 1215.7 | [4470.3] | | |
| | | | | Si IV | 1398.3 | 5108.1 | | |
| | | | | C IV | 1549.5 | 5642.9 | | |
| | | | | C III] | 1908.7 | 6969.9 | | |
| IVS B0423+297 | 04 26 30.2261 | +29 52 22.933 | GN | C IV | 1549.5 | 5087.9 | $2.291 \pm 0.007$ | |
| | | | | C III] | 1908.7 | 6295.6 | | |
| IVS B0431−642 | 04 31 28.1028 | −64 06 31.366 | NTT | Si IV | 1398.3 | 4094.3 | $1.9263 \pm 0.0023$ | |
| | | | | C IV | 1549.5 | 4527.8 | | |

| Source | RA (J2000)[a] | Dec (J2000)[a] | Telescope[b] | Line | Rest $\lambda$ | Obs $\lambda$[c] | $z$ | Note[d] |
|---|---|---|---|---|---|---|---|---|
| | | | | C III] | 1908.7 | 5579.3 | | |
| | | | | Mg II | 2799.9 | 8209.5 | | |
| IVS B0511−203 | 05 13 42.8583 | −20 16 11.487 | NTT | C IV | 1549.5 | 3861.4 | $1.487 \pm 0.004$ | |
| | | | | C III] | 1908.7 | 4732.5 | | |
| | | | | Mg II | 2799.9 | 6974.2 | | |
| IVS B0521−403 | 05 22 54.1151 | −40 20 30.941 | NTT | C III] | 1908.7 | 4549.7 | $1.380 \pm 0.004$ | * |
| | | | | Mg II | 2799.9 | 6652.6 | | |
| IVS B0530−484 | 05 31 58.6151 | −48 27 35.972 | NTT | Mg II | 2799.9 | 5073.2 | $0.8116 \pm 0.0003$ | |
| | | | | [O II] | 3726.8 | 6750.4 | | |
| IVS B0538−562 | 05 39 42.3928 | −56 12 49.967 | NTT | C III] | 1908.7 | 4236.8 | $1.224 \pm 0.004$ | |
| | | | | Mg II | 2799.9 | 6239.5 | | |
| IVS B0550−453 | 05 52 10.0402 | −45 22 24.671 | NTT | Mg II | 2799.9 | 5039.0 | $0.8011 \pm 0.0004$ | * |
| | | | | [Ne V] | 3425.5 | 6165.5 | | |
| | | | | [O II] | 3726.8 | 6710.6 | | |
| | | | | [Ne III] | 3868.7 | 6971.9 | | |
| | | | | H$\epsilon$ | 3970.1 | 7144.7 | | |

| Source | RA (J2000)[a] | Dec (J2000)[a] | Telescope[b] | Line | Rest $\lambda$ | Obs $\lambda$[c] | $z$ | Note[d] |
|---|---|---|---|---|---|---|---|---|
| | | | | H$\delta$ | 4101.7 | 7394.8 | | |
| | | | | H$\gamma$ | 4340.5 | 7825.8 | | |
| | | | | H$\beta$ | 4861.3 | 8757.9 | | |
| | | | | [O III] | 4958.9 | 8930.9 | | |
| | | | | [O III] | 5006.8 | [9015.2] | | |
| IVS B0606−025 | 06 09 28.5116 | −02 30 52.931 | NTT | Mg II | 2799.9 | 4588.7 | $0.6362 \pm 0.0010$ | |
| | | | | [O II] | 3726.8 | 6091.0 | | |
| | | | | [O III] | 4958.9 | 8111.8 | | |
| | | | | [O III] | 5006.8 | 8188.9 | | |
| IVS B0619−468 | 06 21 19.4363 | −46 49 58.438 | NTT | C III] | 1908.7 | 4214.9 | $1.2105 \pm 0.0023$ | |
| | | | | Mg II | 2799.9 | 6195.5 | | |
| IVS B0704−231 | 07 06 33.5180 | −23 11 37.846 | NTT | Mg II | 2799.9 | 4742.3 | $0.6918 \pm 0.0006$ | |
| | | | | [Ne V] | 3345.4 | 5655.5 | | |
| | | | | [Ne V] | 3425.5 | 5790.7 | | |
| | | | | [O II] | 3726.8 | 6300.7 | | |
| | | | | [Ne III] | 3868.7 | 6544.0 | | |

| Source | RA (J2000)[a] | Dec (J2000)[a] | Telescope[b] | Line | Rest $\lambda$ | Obs $\lambda$[c] | $z$ | Note[d] |
|--------|---------------|----------------|--------------|------|--------|--------|-----|---------|
| | | | | [Ne III] | 3967.4 | 6707.2 | | |
| | | | | H$\gamma$ | 4340.5 | 7361.9 | | |
| | | | | [O III] | 4363.2 | 7378.9 | | |
| | | | | H$\beta$ | 4861.3 | 8233.0 | | |
| | | | | [O III] | 4958.9 | 8384.9 | | |
| | | | | [O III] | 5006.8 | 8466.4 | | |
| IVS B0705−412 | 07 06 43.0598 | −41 22 23.785 | NTT | C IV | 1549.5 | 4007.5 | $1.5868 \pm 0.0021$ | |
| | | | | He II | 1640.4 | 4239.8 | | |
| | | | | C III] | 1908.7 | 4930.9 | | |
| | | | | Mg II | 2799.9 | 7259.6 | | |
| IVS B0722−068 | 07 24 29.1437 | −06 58 07.734 | NTT | Mg II | 2799.9 | 4161.5 | $0.4871 \pm 0.0004$ | |
| | | | | H$\gamma$ | 4340.5 | 6458.1 | | |
| | | | | H$\beta$ | 4861.3 | 7231.7 | | |
| | | | | [O III] | 5006.8 | 7444.2 | | |
| IVS B0727−365 | 07 29 05.4122 | −36 39 45.244 | NTT | Mg II | 2799.9 | 4497.6 | $0.6027 \pm 0.0011$ | * |
| | | | | [Ne V] | 3425.5 | 5487.2 | | |

| Source | RA (J2000)[a] | Dec (J2000)[a] | Telescope[b] | Line | Rest $\lambda$ | Obs $\lambda$[c] | $z$ | Note[d] |
|---|---|---|---|---|---|---|---|---|
| | | | | [O II] | 3726.8 | 5962.0 | | |
| | | | | [Ne III] | 3868.7 | 6190.9 | | |
| | | | | [Ne III] | 3967.4 | 6349.5 | | |
| | | | | H$\gamma$ | 4340.5 | 6977.1 | | |
| | | | | H$\beta$ | 4861.3 | 7811.6 | | |
| | | | | [O III] | 4958.9 | 7937.4 | | |
| | | | | [O III] | 5006.8 | 8013.7 | | |
| IVS B0734−044 | 07 36 33.6271 | −04 33 11.859 | NTT | C IV | 1549.5 | 3888.9 | $1.5117 \pm 0.0027$ | |
| | | | | C III] | 1908.7 | 4787.5 | | |
| | | | | Mg II | 2799.9 | 7047.3 | | |
| IVS B0748−069 | 07 50 47.1484 | −07 06 04.025 | NTT | Si IV | 1398.3 | 3865.6 | $1.7624 \pm 0.0023$ | |
| | | | | C IV | 1549.5 | 4284.0 | | |
| | | | | C III] | 1908.7 | 5263.8 | | |
| | | | | Mg II | 2799.9 | [7764.0] | | |
| IVS B0804−267 | 08 06 12.7225 | −26 52 33.308 | NTT | C III] | 1908.7 | 4257.6 | $1.2302 \pm 0.0004$ | |
| | | | | Mg II | 2799.9 | 6243.3 | | |

| Source | RA (J2000)[a] | Dec (J2000)[a] | Telescope[b] | Line | Rest $\lambda$ | Obs $\lambda$[c] | $z$ | Note[d] |
|---|---|---|---|---|---|---|---|---|
| IVS B0812−252 | 08 14 55.1299 | −25 21 43.924 | NTT | Mg II | 2799.9 | 5086.7 | 0.8167 ± 0.0002 | |
| | | | | H$\gamma$ | 4340.5 | [7903.2] | | |
| IVS B0814−029 | 08 17 27.4860 | −03 07 37.311 | NTT | Si IV | 1398.3 | [3897.6] | 1.7748 ± 0.0028 | |
| | | | | C IV | 1549.5 | 4301.3 | | |
| | | | | C III] | 1908.7 | 5286.2 | | |
| | | | | Mg II | 2799.9 | 7781.0 | | |
| IVS B0838−573 | 08 40 09.0201 | −57 32 02.320 | NTT | C IV | 1549.5 | 3764.9 | 1.4264 ± 0.0017 | * |
| | | | | He II | 1640.4 | [3999.0] | | |
| | | | | C III] | 1908.7 | 4627.8 | | |
| | | | | Mg II | 2799.9 | 6789.3 | | |
| IVS B0845−837 | 08 40 09.6134 | −83 54 32.576 | NTT | Mg II | 2799.9 | 4854.9 | 0.7344 ± 0.0005 | |
| | | | | H$\delta$ | 4101.7 | 7115.5 | | |
| | | | | H$\gamma$ | 4340.5 | [7536.3] | | |
| | | | | H$\beta$ | 4861.3 | 8436.7 | | |
| | | | | [O III] | 4958.9 | [8600.5] | | |
| | | | | [O III] | 5006.8 | 8678.8 | | |

| Source | RA (J2000)[a] | Dec (J2000)[a] | Telescope[b] | Line | Rest $\lambda$ | Obs $\lambda$[c] | $z$ | Note[d] |
|---|---|---|---|---|---|---|---|---|
| IVS B0855−716 | 08 55 11.7699 | −71 49 06.457 | NTT | Si IV | 1398.3 | [4000.6] | $1.856 \pm 0.003$ | |
| | | | | C IV | 1549.5 | 4431.3 | | |
| | | | | C III] | 1908.7 | 5438.2 | | |
| | | | | Mg II | 2799.9 | 8001.2 | | |
| IVS B0858−313 | 09 00 44.2942 | −31 31 28.577 | NTT | Mg II | 2799.9 | 4064.7 | $0.4528 \pm 0.0008$ | * |
| | | | | H$\beta$ | 4861.3 | 7070.5 | | |
| | | | | [O III] | 5006.8 | 7271.2 | | |
| IVS B0905−202 | 09 07 54.0404 | −20 26 49.475 | NTT | Mg II | 2799.9 | 5337.4 | $0.906 \pm 0.003$ | S |
| IVS B0917−354 | 09 19 06.8892 | −35 38 27.040 | NTT | Mg II | 2799.9 | 3861.0 | $0.3823 \pm 0.0013$ | * |
| | | | | H$\delta$ | 4101.7 | 5670.2 | | |
| | | | | H$\gamma$ | 4340.5 | 6000.4 | | |
| | | | | H$\beta$ | 4861.3 | 6735.1 | | |
| | | | | [O III] | 5006.8 | [6941.9] | | |
| IVS B0922−202 | 09 25 11.9473 | −20 27 35.610 | NTT | C III] | 1908.7 | 3975.0 | $1.0844 \pm 0.0019$ | |
| | | | | Mg II | 2799.9 | 5841.3 | | |
| IVS B0937−187 | 09 39 21.2577 | −18 58 05.738 | NTT | Mg II | 2799.9 | 4532.4 | $0.6207 \pm 0.0014$ | |

| Source | RA (J2000)[a] | Dec (J2000)[a] | Telescope[b] | Line | Rest $\lambda$ | Obs $\lambda$[c] | $z$ | Note[d] |
|---|---|---|---|---|---|---|---|---|
| | | | | [Ne v] | 3425.5 | [5559.1] | | |
| | | | | H$\gamma$ | 4340.5 | 7051.4 | | |
| | | | | H$\beta$ | 4861.3 | 7893.5 | | |
| | | | | [O iii] | 4958.9 | 8022.0 | | |
| | | | | [O iii] | 5006.8 | 8104.4 | | |
| IVS B1006−093 | 10 08 43.8653 | −09 33 23.359 | NTT | C iii] | 1908.7 | 4247.5 | $1.2255 \pm 0.0003$ | * |
| | | | | Mg ii | 2799.9 | 6231.9 | | |
| IVS B1020−045 | 10 22 44.1124 | −04 43 36.451 | NTT | Ly$\beta$ | 1025.7 | [3730.5] | $2.624 \pm 0.005$ | |
| | | | | Ly$\alpha$ | 1215.7 | 4399.8 | | |
| | | | | N v | 1240.1 | [4483.9] | | |
| | | | | Si iv | 1398.3 | 5086.0 | | |
| | | | | C iv | 1549.5 | 5605.7 | | |
| | | | | Si iii] | 1890.0 | 6845.5 | | |
| | | | | C iii] | 1908.7 | [6924.0] | | |
| IVS B1044−512 | 10 46 38.9030 | −51 30 11.063 | NTT | Ly$\alpha$ | 1215.7 | 4508.6 | $2.698 \pm 0.006$ | |
| | | | | Si iv | 1398.3 | [5194.1] | | |

| Source | RA (J2000)[a] | Dec (J2000)[a] | Telescope[b] | Line | Rest $\lambda$ | Obs $\lambda$[c] | $z$ | Note[d] |
|---|---|---|---|---|---|---|---|---|
| | | | | C IV | 1549.5 | 5723.6 | | |
| | | | | C III] | 1908.7 | 7043.7 | | |
| IVS B1046−463 | 10 48 18.0083 | −46 39 17.458 | NTT | N V | 1240.1 | [3738.9] | $2.0129 \pm 0.0010$ | |
| | | | | Si IV | 1398.3 | 4216.5 | | |
| | | | | C IV | 1549.5 | 4668.8 | | |
| | | | | C III] | 1908.7 | 5746.6 | | |
| | | | | Mg II | 2799.9 | 8433.7 | | |
| IVS B1115−111 | 11 17 34.0736 | −11 23 50.717 | NTT | C III] | 1908.7 | 3753.9 | $0.9667 \pm 0.0006$ | |
| | | | | Mg II | 2799.9 | 5512.7 | | |
| | | | | [O II] | 3726.8 | 7326.5 | | |
| | | | | H$\epsilon$ | 3970.1 | 7805.2 | | |
| | | | | H$\delta$ | 4101.7 | 8063.8 | | |
| | | | | H$\gamma$ | 4340.5 | [8571.0] | | |
| IVS B1145−478 | 11 48 15.6570 | −48 05 38.431 | NTT | C III] | 1908.7 | 4437.8 | $1.328 \pm 0.003$ | |
| | | | | Mg II | 2799.9 | 6527.6 | | |
| IVS B1219−457 | 12 21 46.7644 | −45 59 06.392 | GS | [Ne IV] | 2423.8 | 4971.1 | $1.052 \pm 0.001$ | * |

| Source | RA (J2000)[a] | Dec (J2000)[a] | Telescope[b] | Line | Rest $\lambda$ | Obs $\lambda$[c] | $z$ | Note[d] |
|---|---|---|---|---|---|---|---|---|
| | | | | Mg II | 2799.9 | 5744.9 | | |
| | | | | [Ne v] | 3425.5 | 7033.7 | | |
| IVS B1228−352 | 12 31 16.5421 | −35 33 18.916 | NTT | Mg II | 2799.9 | 4993.7 | 0.7847 ± 0.0006 | |
| | | | | H$\gamma$ | 4340.5 | 7753.8 | | |
| | | | | H$\beta$ | 4861.3 | 8676.7 | | |
| | | | | [O III] | 4958.9 | [8834.8] | | |
| | | | | [O III] | 5006.8 | 8933.0 | | |
| IVS B1229−123 | 12 31 50.2657 | −12 36 37.056 | NTT | Ly$\alpha$ | 1215.7 | 3819.6 | 2.115 ± 0.009 | |
| | | | | C IV | 1549.5 | 4816.8 | | |
| | | | | C III] | 1908.7 | 5926.5 | | |
| | | | | Mg II | 2799.9 | 8696.5 | | |
| IVS B1236−381 | 12 38 52.7315 | −38 25 56.991 | NTT | C IV | 1549.5 | 3870.1 | 1.4974 ± 0.0010 | |
| | | | | He II | 1640.4 | [4099.7] | | |
| | | | | C III] | 1908.7 | 4763.4 | | |
| | | | | Mg II | 2799.9 | 6996.6 | | |
| IVS B1320−385 | 13 23 04.1483 | −38 49 00.671 | GS | C IV | 1549.5 | 4977.5 | 2.2127 ± 0.0005 | |

− 33 −

| Source | RA (J2000)[a] | Dec (J2000)[a] | Telescope[b] | Line | Rest $\lambda$ | Obs $\lambda$[c] | $z$ | Note[d] |
|---|---|---|---|---|---|---|---|---|
| | | | | C III] | 1908.7 | 6132.9 | | |
| IVS B1418−481 | 14 21 38.6465 | −48 20 22.796 | GS | C IV | 1549.5 | 4409.9 | $1.851 \pm 0.005$ | * |
| | | | | C III] | 1908.7 | 5450.2 | | |
| IVS B1520−334 | 15 23 54.0398 | −33 38 45.236 | GS | Mg II | 2799.9 | 4575.5 | $0.6343 \pm 0.0002$ | |
| | | | | [Ne V] | 3425.5 | 5600.2 | | |
| | | | | [Ne III] | 3868.7 | 6323.2 | | |
| | | | | H$\zeta$ | 3889.0 | 6357.6 | | |
| | | | | H$\epsilon$ | 3970.1 | 6485.8 | | |
| | | | | H$\delta$ | 4101.7 | 6705.9 | | |
| | | | | H$\gamma$ | 4340.5 | 7094.1 | | |
| | | | | [O III] | 4363.2 | 7127.5 | | |
| IVS B1531−225 | 15 34 22.8012 | −22 43 57.629 | GS | Mg II | 2799.9 | 5546.9 | $0.981 \pm 0.002$ | S |
| IVS B1540−828 | 15 50 59.1404 | −82 58 06.838 | GS | C IV | 1549.5 | 4243.7 | $1.7383 \pm 0.0006$ | |
| | | | | C III] | 1908.7 | 5226.1 | | |
| | | | | C II] | 2326.9 | 6368.2 | | |
| | | | | [Ne IV] | 2423.8 | 6640.0 | | |

| Source | RA (J2000)[a] | Dec (J2000)[a] | Telescope[b] | Line | Rest $\lambda$ | Obs $\lambda$[c] | $z$ | Note[d] |
|---|---|---|---|---|---|---|---|---|
| IVS B1541−230 | 15 44 14.1780 | −23 12 01.347 | GS | Ly$\alpha$ | 1215.7 | 4462.0 | 2.664 ± 0.004 | |
| | | | | C IV | 1549.5 | 5668.6 | | |
| | | | | C III] | 1908.7 | 6989.8 | | |
| IVS B1545−760 | 15 52 20.4427 | −76 14 46.933 | GS | [O II] | 3726.8 | 4575.9 | 0.2273 ± 0.0002 | |
| | | | | [Ne III] | 3868.7 | [4752.5] | | |
| | | | | H$\gamma$ | 4340.5 | [5338.1] | | |
| | | | | H$\beta$ | 4861.3 | 5965.5 | | |
| | | | | [O III] | 4958.9 | 6085.6 | | |
| | | | | [O III] | 5006.8 | 6144.4 | | |
| IVS B1558−072 | 16 00 56.4765 | −07 22 05.224 | GS | Si IV | 1398.3 | 4197.6 | 2.0016 ± 0.0002 | * |
| | | | | C IV | 1549.5 | 4650.3 | | |
| | | | | He II | 1640.4 | [4921.0] | | |
| | | | | C III] | 1908.7 | 5729.7 | | |
| IVS B1606−742 | 16 12 33.8838 | −74 23 40.137 | GS | Mg II | 2799.9 | 4603.3 | 0.6453 ± 0.0005 | |
| | | | | [Ne V] | 3425.5 | 5640.8 | | |
| | | | | [O II] | 3726.8 | 6131.4 | | |

| Source | RA (J2000)[a] | Dec (J2000)[a] | Telescope[b] | Line | Rest $\lambda$ | Obs $\lambda$[c] | $z$ | Note[d] |
|---|---|---|---|---|---|---|---|---|
| | | | | [Ne III] | 3868.7 | 6365.3 | | |
| | | | | H$\gamma$ | 4340.5 | [7155.8] | | |
| IVS B1621+058 | 16 24 07.7338 | +05 43 24.244 | GS | C III] | 1908.7 | 4821.7 | $1.5273 \pm 0.0012$ | |
| | | | | Mg II | 2799.9 | 7079.5 | | |
| IVS B1628−771 | 16 36 02.9612 | −77 13 34.420 | GS | [O II] | 3726.8 | 5072.3 | $0.3605 \pm 0.0001$ | |
| | | | | [Ne III] | 3868.7 | 5263.6 | | |
| | | | | H$\beta$ | 4861.3 | 6612.8 | | |
| | | | | [O III] | 4958.9 | 6746.2 | | |
| | | | | [O III] | 5006.8 | 6811.3 | | |
| IVS B1643+007 | 16 46 06.9722 | +00 42 27.279 | GS | C III] | 1908.7 | 4530.0 | $1.3726 \pm 0.0008$ | |
| | | | | Mg II | 2799.9 | 6640.9 | | |
| IVS B1700−577 | 17 04 18.0630 | −57 51 01.604 | GS | [Ne V] | 3425.5 | [4738.9] | $0.3833 \pm 0.0001$ | |
| | | | | [O II] | 3726.8 | 5154.0 | | |
| | | | | [Ne III] | 3868.7 | 5352.7 | | |
| | | | | [Ne III] | 3967.4 | [5488.7] | | |
| | | | | H$\delta$ | 4101.7 | [5666.0] | | |

| Source | RA (J2000)[a] | Dec (J2000)[a] | Telescope[b] | Line | Rest $\lambda$ | Obs $\lambda$[c] | $z$ | Note[d] |
|---|---|---|---|---|---|---|---|---|
| | | | | H$\gamma$ | 4340.5 | [6017.0] | | |
| | | | | [O III] | 4363.2 | 6034.0 | | |
| | | | | H$\beta$ | 4861.3 | 6726.3 | | |
| | | | | [O III] | 4958.9 | 6859.7 | | |
| | | | | [O III] | 5006.8 | 6927.0 | | |
| IVS B1701−132 | 17 04 05.0868 | −13 16 34.234 | GS | Ly$\alpha$ | 1215.7 | [4254.8] | $2.4916 \pm 0.0010$ | |
| | | | | C IV | 1549.5 | 5408.7 | | |
| | | | | C III] | 1908.7 | 6666.3 | | |
| IVS B1712+530 | 17 13 49.8706 | +53 00 32.790 | GN | Si IV | 1398.3 | 4529.4 | $2.233 \pm 0.003$ | * |
| | | | | C IV | 1549.5 | 4996.9 | | |
| | | | | Si III] | 1890.0 | 6116.1 | | |
| | | | | C III] | 1908.7 | 6166.3 | | |
| IVS B1717−618 | 17 21 39.0167 | −61 54 43.020 | GS | Mg II | 2799.9 | 4643.4 | $0.6579 \pm 0.0002$ | |
| | | | | [Ne V] | 3425.5 | 5679.4 | | |
| | | | | [O II] | 3726.8 | 6177.1 | | |
| | | | | [Ne III] | 3868.7 | 6412.8 | | |

| Source | RA (J2000)[a] | Dec (J2000)[a] | Telescope[b] | Line | Rest $\lambda$ | Obs $\lambda$[c] | $z$ | Note[d] |
|---|---|---|---|---|---|---|---|---|
| | | | | H$\gamma$ | 4340.5 | [7216.1] | | |
| IVS B1800−709 | 18 06 23.6852 | −70 58 28.762 | GS | C III] | 1908.7 | [4398.4] | $1.303 \pm 0.001$ | S |
| | | | | Mg II | 2799.9 | 6449.3 | | |
| IVS B1810−745 | 18 16 40.0762 | −74 29 07.343 | GS | C IV | 1549.5 | 5035.6 | $2.256 \pm 0.006$ | * |
| | | | | C III] | 1908.7 | 6224.7 | | |
| IVS B1813−497 | 18 16 55.9942 | −49 43 44.671 | GS | C IV | 1549.5 | 4176.9 | $1.697 \pm 0.002$ | * |
| | | | | C III] | 1908.7 | 5151.2 | | |
| IVS B1905−252 | 19 08 48.4392 | −25 09 29.373 | GS | C IV | 1549.5 | 4095.7 | $1.6434 \pm 0.0003$ | * |
| | | | | C III] | 1908.7 | 5045.9 | | |
| IVS B1905−794 | 19 13 33.2602 | −79 20 14.810 | GS | Mg II | 2799.9 | 5720.7 | $1.043 \pm 0.001$ | S |
| IVS B1939−316 | 19 42 40.9140 | −31 30 14.609 | GS | C III] | 1908.7 | 4927.5 | $1.5818 \pm 0.0002$ | |
| | | | | Mg II | 2799.9 | 7229.2 | | |
| IVS B1941−372 | 19 44 57.8355 | −37 07 09.874 | GS | C IV | 1549.5 | 4396.2 | $1.840 \pm 0.002$ | |
| | | | | He II | 1640.4 | [4669.8] | | |
| | | | | C III] | 1908.7 | 5424.5 | | |
| IVS B1957−423 | 20 00 56.7918 | −42 14 46.269 | GS | Mg II | 2799.9 | 5575.9 | $0.9889 \pm 0.0025$ | * |

| Source | RA (J2000)[a] | Dec (J2000)[a] | Telescope[b] | Line | Rest $\lambda$ | Obs $\lambda$[c] | $z$ | Note[d] |
|---|---|---|---|---|---|---|---|---|
| | | | | [Ne V] | 3425.5 | 6804.5 | | |
| IVS B2029−215 | 20 32 54.3154 | −21 25 15.756 | GS | Mg II | 2799.9 | 5241.0 | $0.8722 \pm 0.0004$ | * |
| | | | | [O II] | 3726.8 | 6978.9 | | |
| IVS B2036−034 | 20 39 09.9852 | −03 17 14.411 | GS | C III] | 1908.7 | [4872.3] | $1.551 \pm 0.002$ | S * |
| | | | | Mg II | 2799.9 | 7142.6 | | |
| IVS B2039−078 | 20 42 04.9182 | −07 42 37.845 | GS | Mg II | 2799.9 | 5727.0 | $1.045 \pm 0.001$ | S |
| IVS B2044−188 | 20 47 37.6552 | −18 41 41.352 | GS | Ly$\beta$ | 1025.7 | [4128.5] | $3.002 \pm 0.004$ | * |
| | | | | Ly$\alpha$ | 1215.7 | 4873.8 | | |
| | | | | N V | 1240.1 | 4949.1 | | |
| | | | | Si II | 1265.2 | [5070.2] | | |
| | | | | O I/Si II | 1305.4 | 5235.7 | | |
| | | | | Si IV | 1398.3 | 5595.6 | | |
| | | | | C IV | 1549.5 | 6187.7 | | |
| IVS B2048−557 | 20 52 13.6851 | −55 33 10.043 | GS | C III] | 1908.7 | [4771.4] | $1.502 \pm 0.002$ | S |
| | | | | Mg II | 2799.9 | 7005.6 | | |
| IVS B2051−204 | 20 54 22.0724 | −20 16 16.819 | GS | Mg II | 2799.9 | 5577.8 | $0.992 \pm 0.001$ | S |

| Source | RA (J2000)[a] | Dec (J2000)[a] | Telescope[b] | Line | Rest $\lambda$ | Obs $\lambda$[c] | $z$ | Note[d] |
|--------|---------------|----------------|--------------|------|----------------|------------------|-----|---------|
| IVS B2107−553 | 21 11 12.1527 | −55 08 06.172 | GS | Ly$\alpha$ | 1215.7 | [4141.2] | $2.3965 \pm 0.0005$ | * |
| | | | | Si IV | 1398.3 | [4756.3] | | |
| | | | | C IV | 1549.5 | 5262.2 | | |
| | | | | C III] | 1908.7 | 6483.8 | | |
| IVS B2118−037 | 21 20 48.4735 | −03 30 28.929 | GS | C IV | 1549.5 | 4207.2 | $1.7169 \pm 0.0017$ | |
| | | | | C III] | 1908.7 | 5188.9 | | |
| IVS B2131−340 | 21 34 50.8245 | −33 51 14.750 | GS | C IV | 1549.5 | 4206.0 | $1.7150 \pm 0.0006$ | |
| | | | | C III] | 1908.7 | 5183.2 | | |
| IVS B2131−690 | 21 35 49.8826 | −68 48 33.925 | GS | C III] | 1908.7 | 4688.6 | $1.464 \pm 0.008$ | * |
| | | | | Mg II | 2799.9 | 6922.2 | | |
| IVS B2132−638 | 21 36 22.0749 | −63 35 51.038 | GS | C III] | 1908.7 | 4835.6 | $1.537 \pm 0.004$ | * |
| | | | | Mg II | 2799.9 | 7113.7 | | |
| IVS B2137−644 | 21 41 46.4411 | −64 11 14.476 | GS | [Ne V] | 3425.5 | [6704.1] | $0.959 \pm 0.003$ | S |
| | | | | Mg II | 2799.9 | 5484.8 | | |
| IVS B2155−475 | 21 58 40.2025 | −47 19 26.552 | GS | C IV | 1549.5 | 4238.6 | $1.7371 \pm 0.0017$ | * |
| | | | | C III] | 1908.7 | 5227.5 | | |

| Source | RA (J2000)[a] | Dec (J2000)[a] | Telescope[b] | Line | Rest $\lambda$ | Obs $\lambda$[c] | $z$ | Note[d] |
|---|---|---|---|---|---|---|---|---|
| IVS B2157−255 | 21 59 50.7922 | −25 16 52.176 | GS | Mg II | 2799.9 | 4693.7 | $0.6757 \pm 0.0002$ | |
| | | | | [Ne v] | 3425.5 | 5738.7 | | |
| | | | | [O II] | 3726.8 | 6245.1 | | |
| | | | | [Ne III] | 3868.7 | 6481.8 | | |
| IVS B2211+069 | 22 14 08.8615 | +07 11 42.393 | GN | [O II] | 3726.8 | 5373.0 | $0.4412 \pm 0.0004$ | |
| | | | | [O III] | 4958.9 | 7143.3 | | |
| | | | | [O III] | 5006.8 | 7212.7 | | |
| | | | | Mg b | 5175.0 | 7462.9 | | |
| IVS B2225−694 | 22 29 00.1767 | −69 10 30.274 | GS | Mg II | 2799.9 | 5340.0 | $0.907 \pm 0.001$ | S |
| IVS B2226−634 | 22 30 10.3381 | −63 10 43.299 | GS | Ly$\alpha$ | 1215.7 | 4111.8 | $2.3841 \pm 0.0010$ | |
| | | | | N v | 1240.1 | [4191.8] | | |
| | | | | Si IV | 1398.3 | [4730.4] | | |
| | | | | C IV | 1549.5 | 5245.9 | | |
| | | | | C III] | 1908.7 | 6459.9 | | |
| IVS B2235−556 | 22 39 06.0344 | −55 25 59.410 | GS | C IV | 1549.5 | 4599.6 | $1.9706 \pm 0.0022$ | |
| | | | | He II | 1640.4 | [4874.0] | | |

| Source | RA (J2000)[a] | Dec (J2000)[a] | Telescope[b] | Line | Rest $\lambda$ | Obs $\lambda$[c] | $z$ | Note[d] |
|---|---|---|---|---|---|---|---|---|
| | | | | C III] | 1908.7 | 5674.1 | | |
| IVS B2243−081 | 22 45 49.0038 | −07 55 19.381 | GS | C IV | 1549.5 | 4347.3 | $1.8069 \pm 0.0005$ | |
| | | | | Al III | 1857.4 | 5214.0 | | |
| | | | | Si III] | 1890.0 | 5307.0 | | |
| | | | | C III] | 1908.7 | 5357.4 | | |
| IVS B2250+323 | 22 53 12.4997 | +32 36 04.326 | GN | [O II] | 3726.8 | 4686.3 | $0.2580 \pm 0.0004$ | |
| | | | | [Ne III] | 3868.7 | 4865.9 | | |
| | | | | [Ne III] | 3967.4 | 4990.5 | | |
| | | | | H$\epsilon$ | 3971.2 | 5008.2 | | |
| | | | | H$\delta$ | 4101.7 | 5156.4 | | |
| | | | | H$\gamma$ | 4340.5 | 5462.4 | | |
| | | | | [O III] | 4363.2 | 5489.0 | | |
| | | | | H$\beta$ | 4861.3 | 6114.6 | | |
| | | | | [O III] | 4958.9 | 6233.5 | | |
| | | | | [O III] | 5006.8 | 6294.9 | | |
| IVS B2255−705 | 22 58 31.5632 | −70 14 39.728 | NTT | C III] | 1908.7 | 3956.9 | $1.077 \pm 0.004$ | |

| Source | RA (J2000)[a] | Dec (J2000)[a] | Telescope[b] | Line | Rest $\lambda$ | Obs $\lambda$[c] | $z$ | Note[d] |
|---|---|---|---|---|---|---|---|---|
| | | | | Mg II | 2799.9 | 5826.9 | | |
| IVS B2300+386 | 23 03 04.0658 | +38 53 48.365 | GN | C IV | 1549.5 | 4253.4 | $1.7462 \pm 0.0013$ | * |
| | | | | C III] | 1908.7 | 5244.2 | | |
| IVS B2311+276 | 23 13 54.8539 | +27 57 04.514 | GN | Ly$\alpha$ | 1215.7 | [4651.2] | $2.801 \pm 0.006$ | |
| | | | | Si IV | 1398.3 | [5293.9] | | |
| | | | | C IV | 1549.5 | 5880.7 | | |
| | | | | C III] | 1908.7 | 7267.4 | | |
| IVS B2313−182 | 23 15 48.1321 | −18 00 40.251 | NTT | [O II] | 3726.8 | 5270.3 | $0.4126 \pm 0.0013$ | * |
| | | | | [O III] | 4958.9 | 7013.4 | | |
| | | | | [O III] | 5006.8 | 7080.2 | | |
| | | | | Ca K | 3934.8 | 5559.9 | | |
| | | | | Ca H | 3969.6 | 5586.6 | | |
| IVS B2317−372 | 23 19 53.5643 | −36 56 28.265 | NTT | Mg II | 2799.9 | 4140.0 | $0.4805 \pm 0.0005$ | |
| | | | | [O II] | 3726.8 | 5519.4 | | |
| | | | | H$\beta$ | 4861.3 | 7197.7 | | |
| | | | | [O III] | 4958.9 | 7344.5 | | |

| Source | RA (J2000)[a] | Dec (J2000)[a] | Telescope[b] | Line | Rest $\lambda$ | Obs $\lambda$[c] | $z$ | Note[d] |
|--------|---------------|----------------|--------------|------|----------------|------------------|-----|---------|
|  |  |  |  | [O III] | 5006.8 | 7416.1 |  |  |

[a]ICRF2 radio position

[b]Telescope abbreviations: GN = Gemini North, GS = Gemini South, NTT = ESO New Technology Telescope

[c]Square brackets indicate the line was not used for redshift determination

[d]"S" denotes a single-line redshift; * indicates a note in Section 3.1

Table 2.   Probable BL Lac objects

| Source | RA (J2000)[a] | Dec (J2000)[a] | Telescope |
|---|---|---|---|
| IVS B0019−261 | 00 21 32.5495 | −25 50 49.387 | GS |
| IVS B0138−381 | 01 40 36.8208 | −37 56 45.798 | NTT |
| IVS B0141+268 | 01 44 33.5538 | +27 05 03.119 | GN |
| IVS B0142−590 | 01 43 47.4305 | −58 45 51.375 | NTT |
| IVS B0258−170 | 03 01 16.6227 | −16 52 45.101 | GS |
| IVS B0839−357 | 08 41 21.6331 | −35 55 05.967 | NTT |
| IVS B0843−547 | 08 45 02.4822 | −54 58 08.545 | NTT |
| IVS B1029−854 | 10 26 34.3753 | −85 43 14.295 | NTT |
| IVS B1101−536 | 11 03 52.2217 | −53 57 00.697 | NTT |
| IVS B1304−427 | 13 07 37.9835 | −42 59 38.987 | NTT |
| IVS B1711−670 | 17 16 22.3365 | −67 06 24.119 | GS |
| IVS B2257+313 | 23 00 22.8374 | +31 37 04.405 | GN |
| IVS B2308+018 | 23 11 01.2929 | +02 05 05.327 | GN |
| IVS B2310+062 | 23 13 07.7015 | +06 28 38.847 | GN |

[a]ICRF2 radio position

Table 3.   Objects with low signal-to-noise spectra

| Source | RA (J2000)[a] | Dec (J2000)[a] | Telescope |
|---|---|---|---|
| IVS B0002−350 | 00 05 05.9251 | −34 45 49.656 | NTT |
| IVS B0054−451 | 00 56 45.8532 | −44 51 02.141 | NTT, GS |
| IVS B0059−628 | 01 01 14.9683 | −62 33 06.339 | GS |
| IVS B0131−795 | 01 31 48.8223 | −79 16 17.486 | NTT |
| IVS B0253−754 | 02 53 10.8817 | −75 13 16.686 | GS |
| IVS B0255+203 | 02 58 07.3094 | +20 30 01.580 | GN |
| IVS B0302−063 | 03 05 00.5637 | −06 07 41.500 | GS |
| IVS B0312−512 | 03 14 25.7036 | −51 04 31.562 | GS |
| IVS B0545+221 | 05 48 03.8896 | +22 10 35.593 | GN |
| IVS B1004−500 | 10 06 14.0093 | −50 18 13.471 | GS |
| IVS B1151−324 | 11 54 06.1664 | −32 42 42.983 | NTT |
| IVS B1517−385 | 15 20 40.4844 | −38 45 32.412 | GS |
| IVS B1532−473 | 15 35 52.2412 | −47 30 22.978 | GS |
| IVS B1650−157 | 16 53 34.2064 | −15 51 29.888 | GS |
| IVS B1653−612 | 16 57 49.0093 | −61 21 37.859 | GS |
| IVS B1732−598 | 17 36 30.8521 | −59 51 58.749 | GS |
| IVS B1830−589 | 18 34 27.4734 | −58 56 36.272 | GS |
| IVS B1914−412 | 19 18 16.0574 | −41 11 31.055 | GS |
| IVS B1935−452 | 19 39 30.0868 | −45 10 01.273 | GS |
| IVS B2009+795 | 20 07 13.1610 | +79 42 11.336 | GN |
| IVS B2025−538 | 20 29 35.0553 | −53 39 07.295 | GS |

Table 3—Continued

| Source | RA (J2000)[a] | Dec (J2000)[a] | Telescope |
|---|---|---|---|
| IVS B2210+277 | 22 12 39.1030 | +27 59 38.449 | GN |

[a]ICRF2 radio position

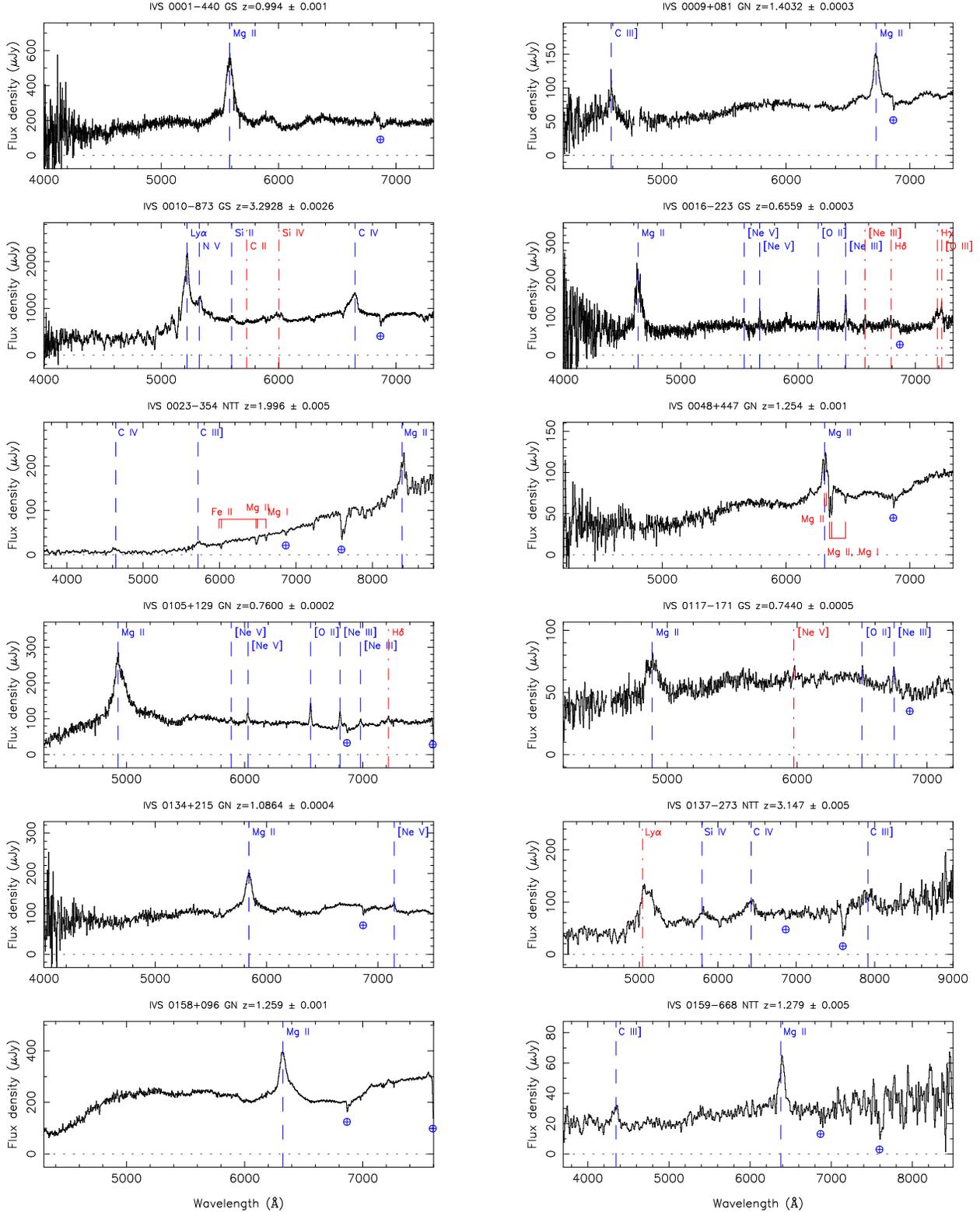

Fig. 1.— Optical spectra for 112 emission-line IVS targets. Dashed lines (blue) indicate emission lines used for redshift determination; dot-dashed lines (red) indicate lines detected at lower signal-to-noise or blended. Short gaps in Gemini spectra are due to physical gaps between the GMOS CCDs. Occasional larger gaps arise from calibration problems or strong

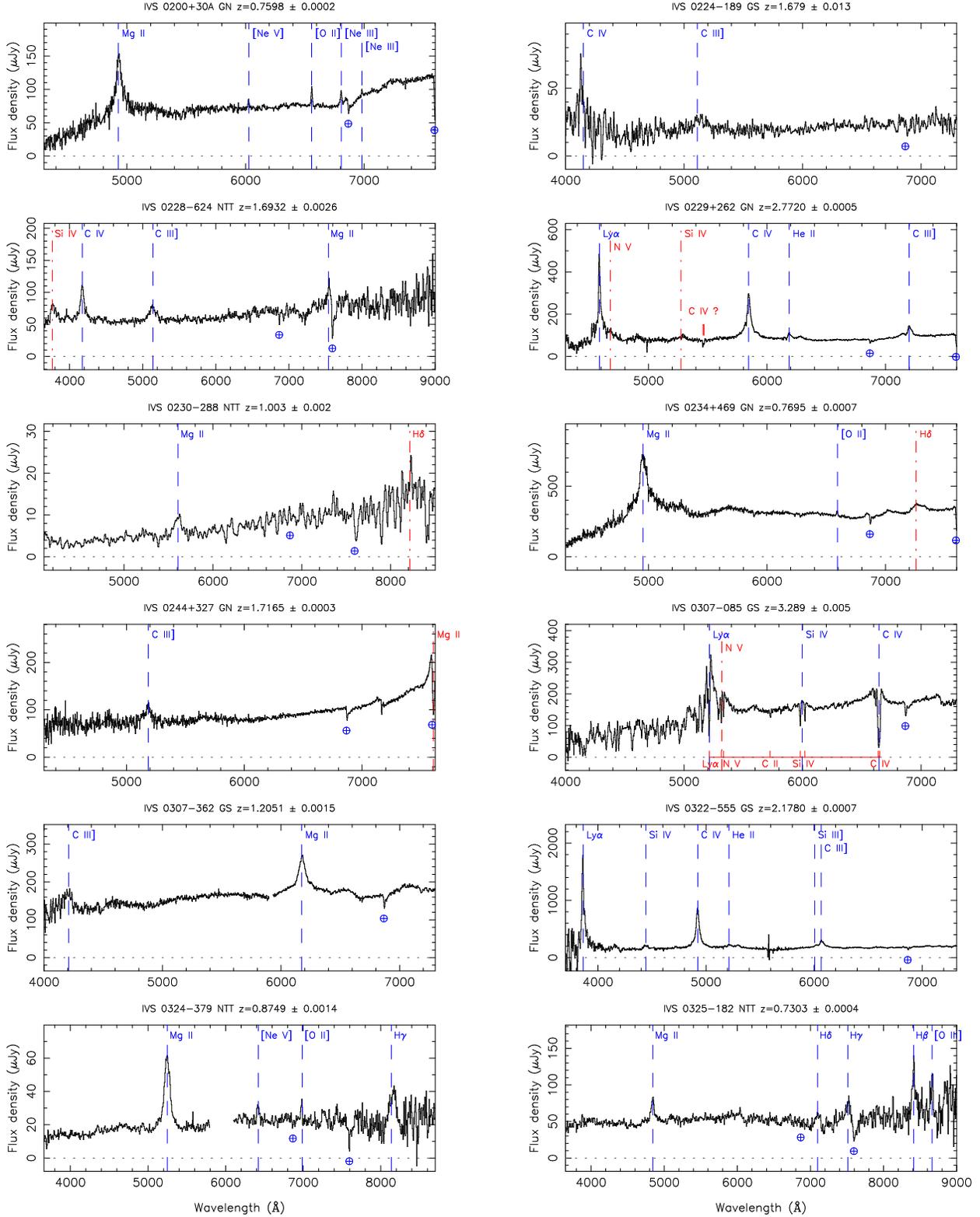

Fig. 1 (continued).—

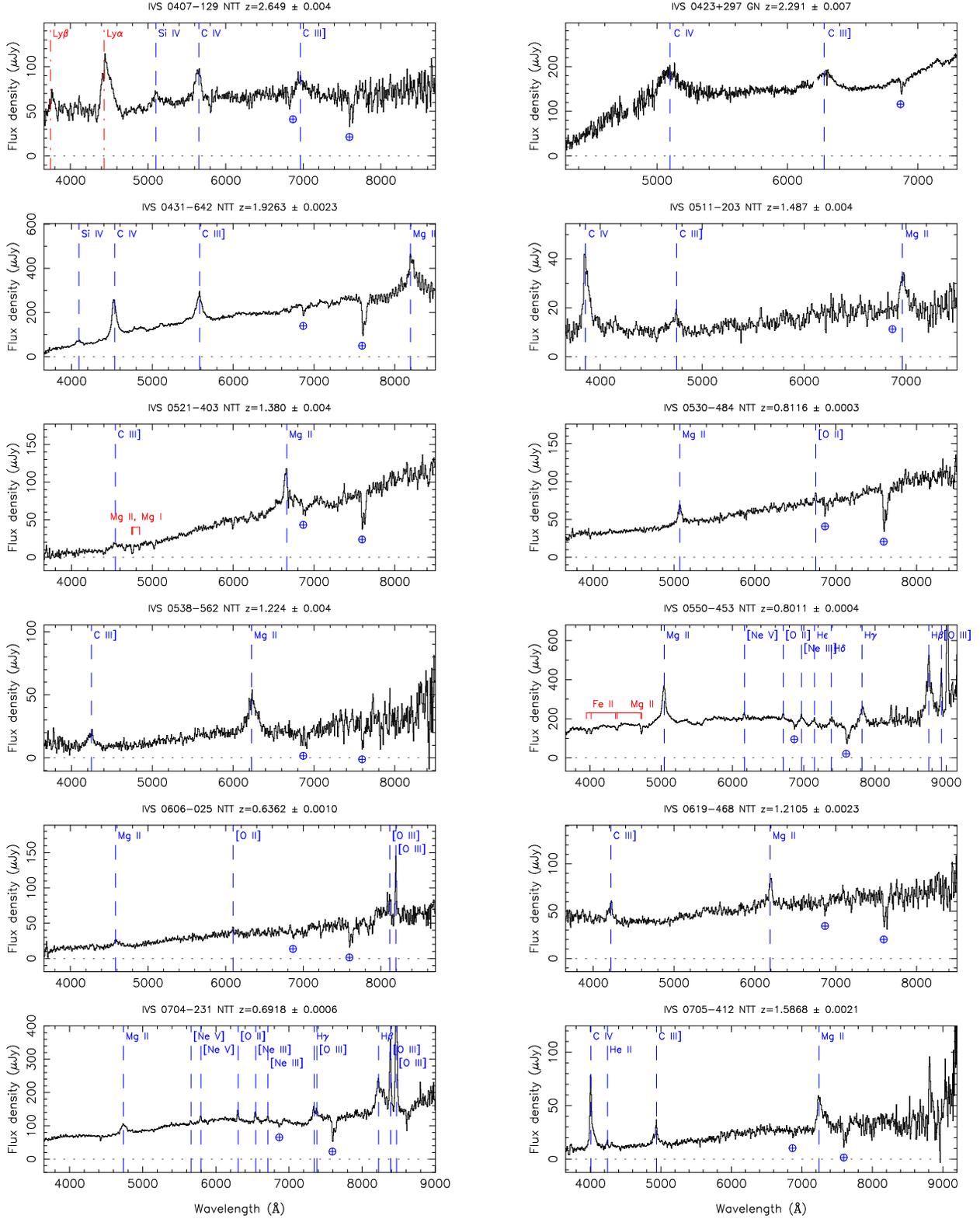

Fig. 1 (continued).—

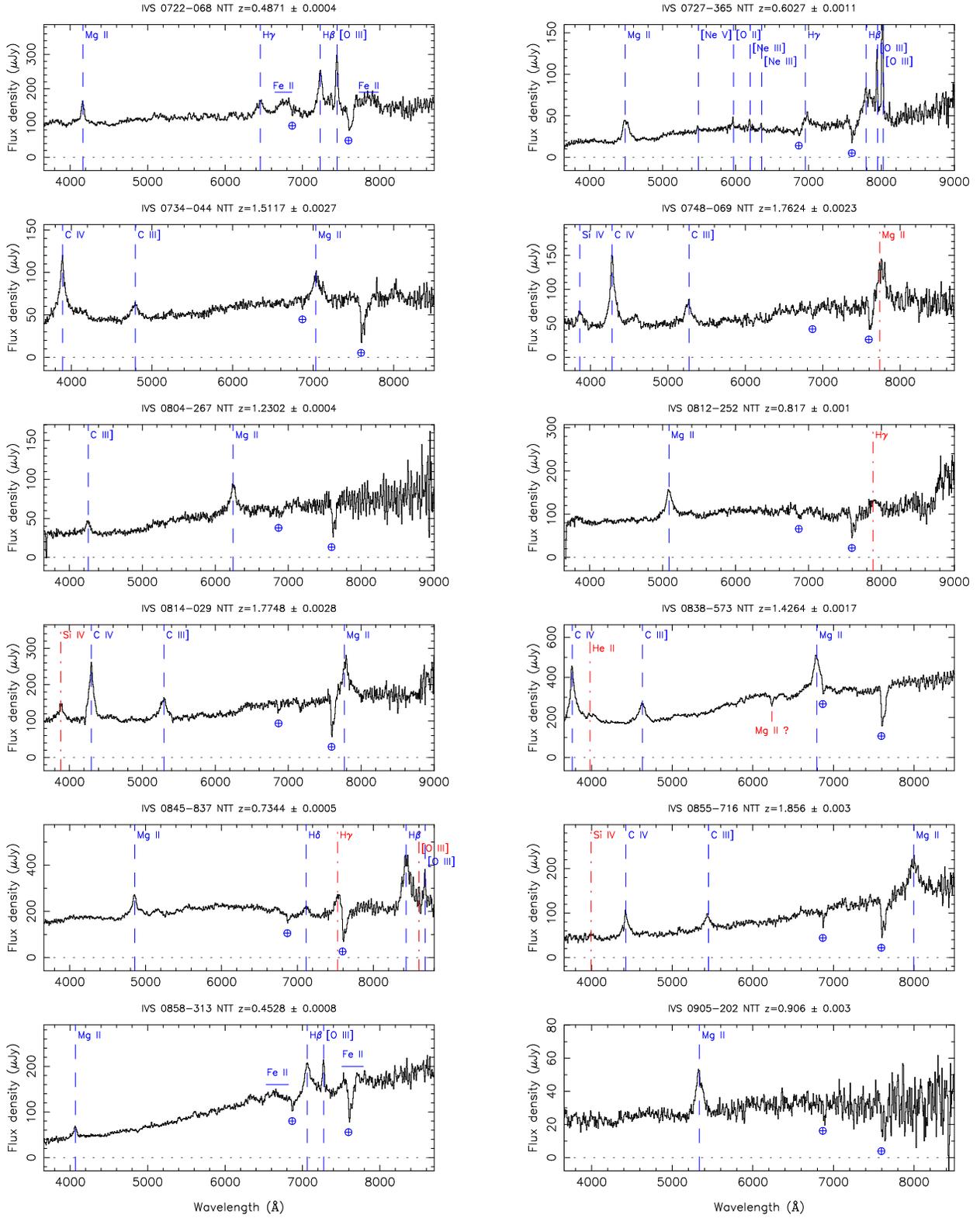

Fig. 1 (continued).—

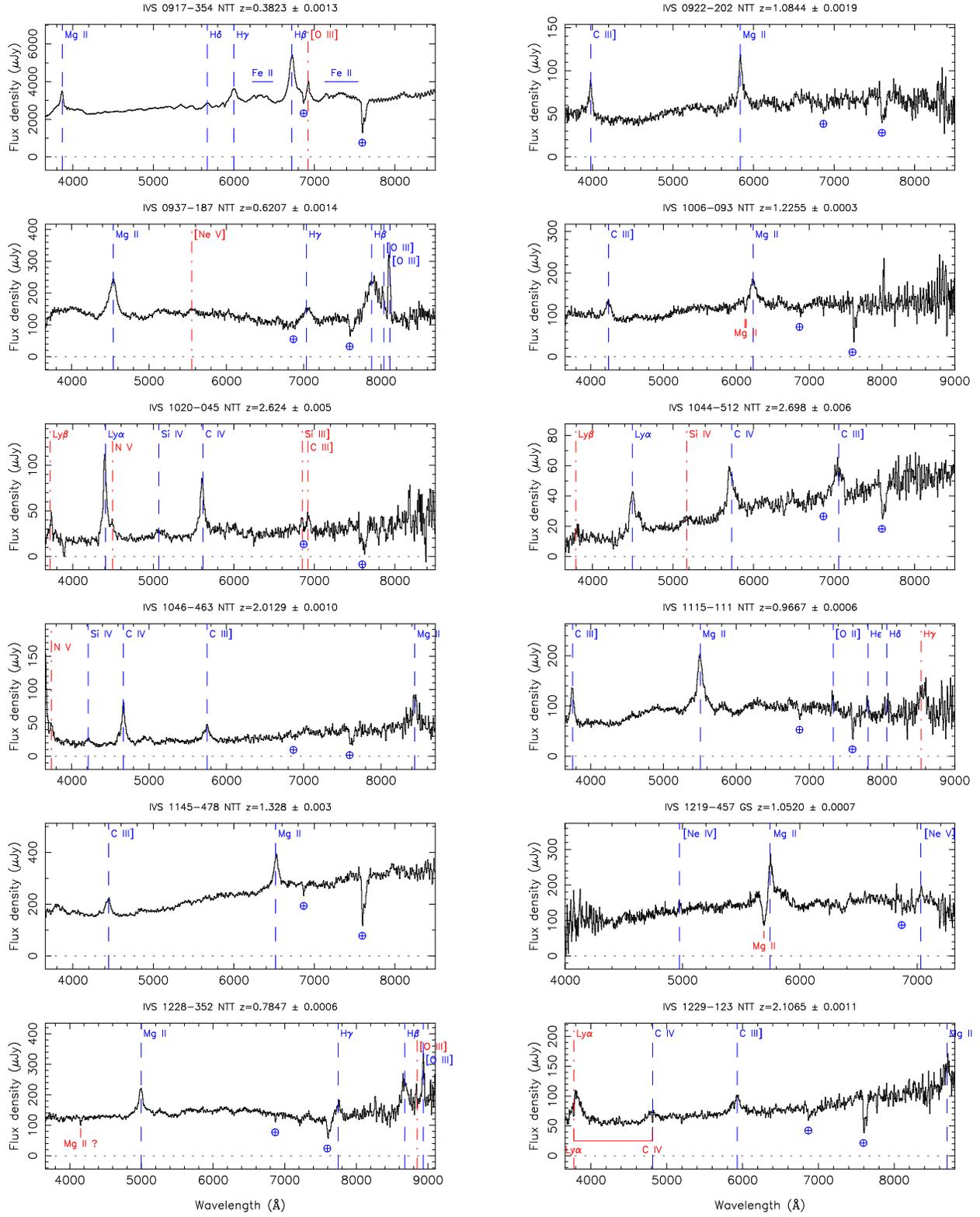

Fig. 1 (continued).—

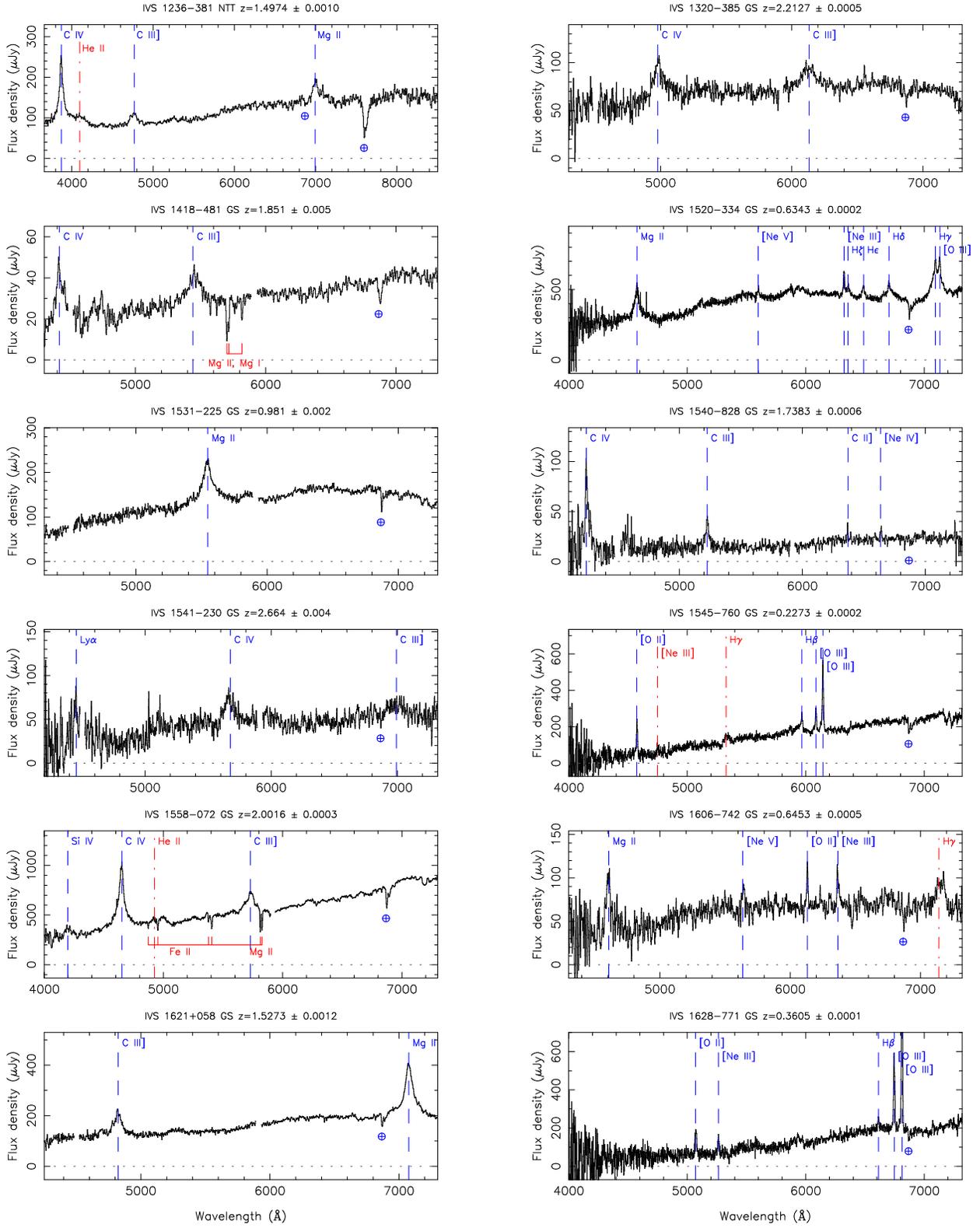

Fig. 1 (continued).—

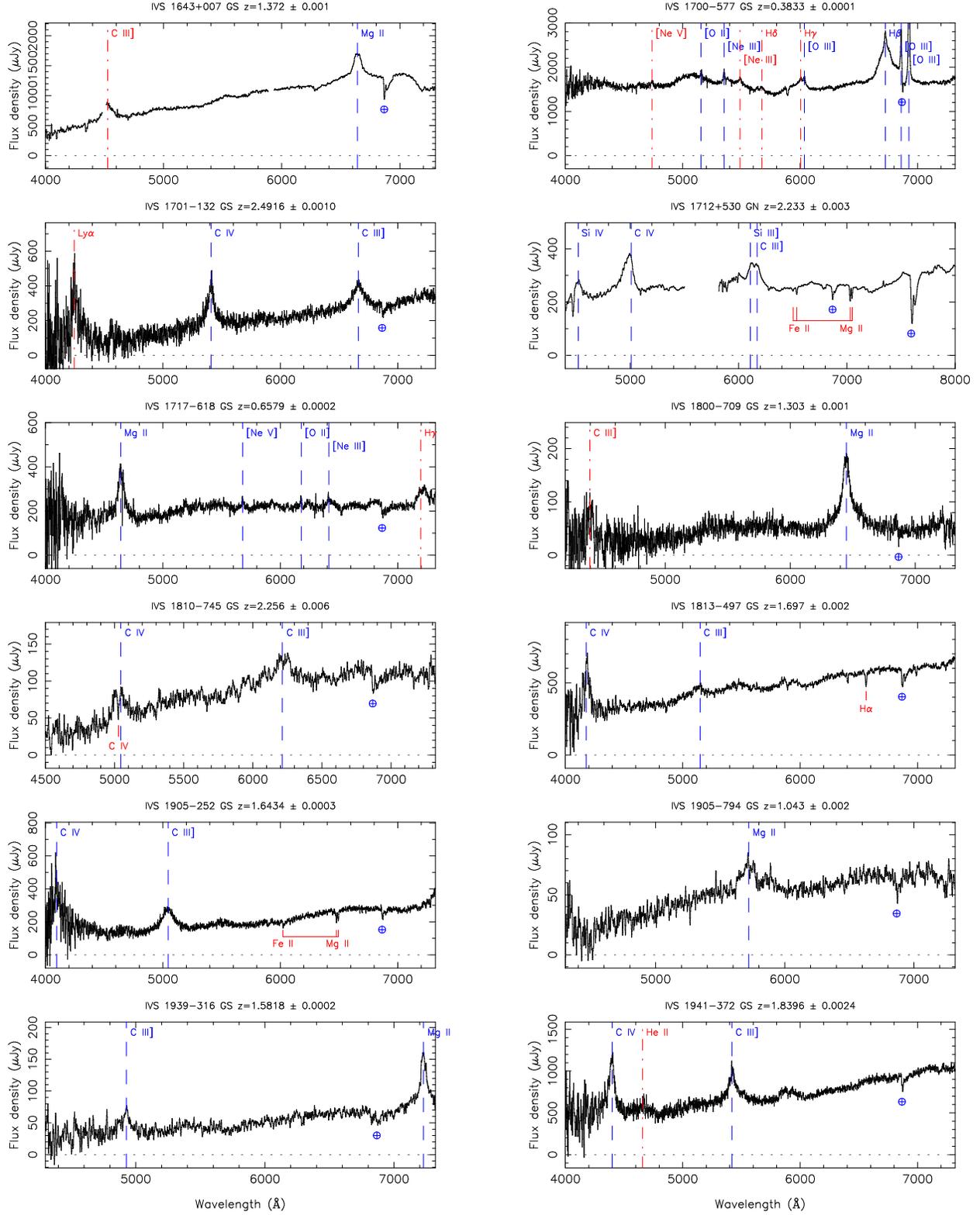

Fig. 1 (continued).—

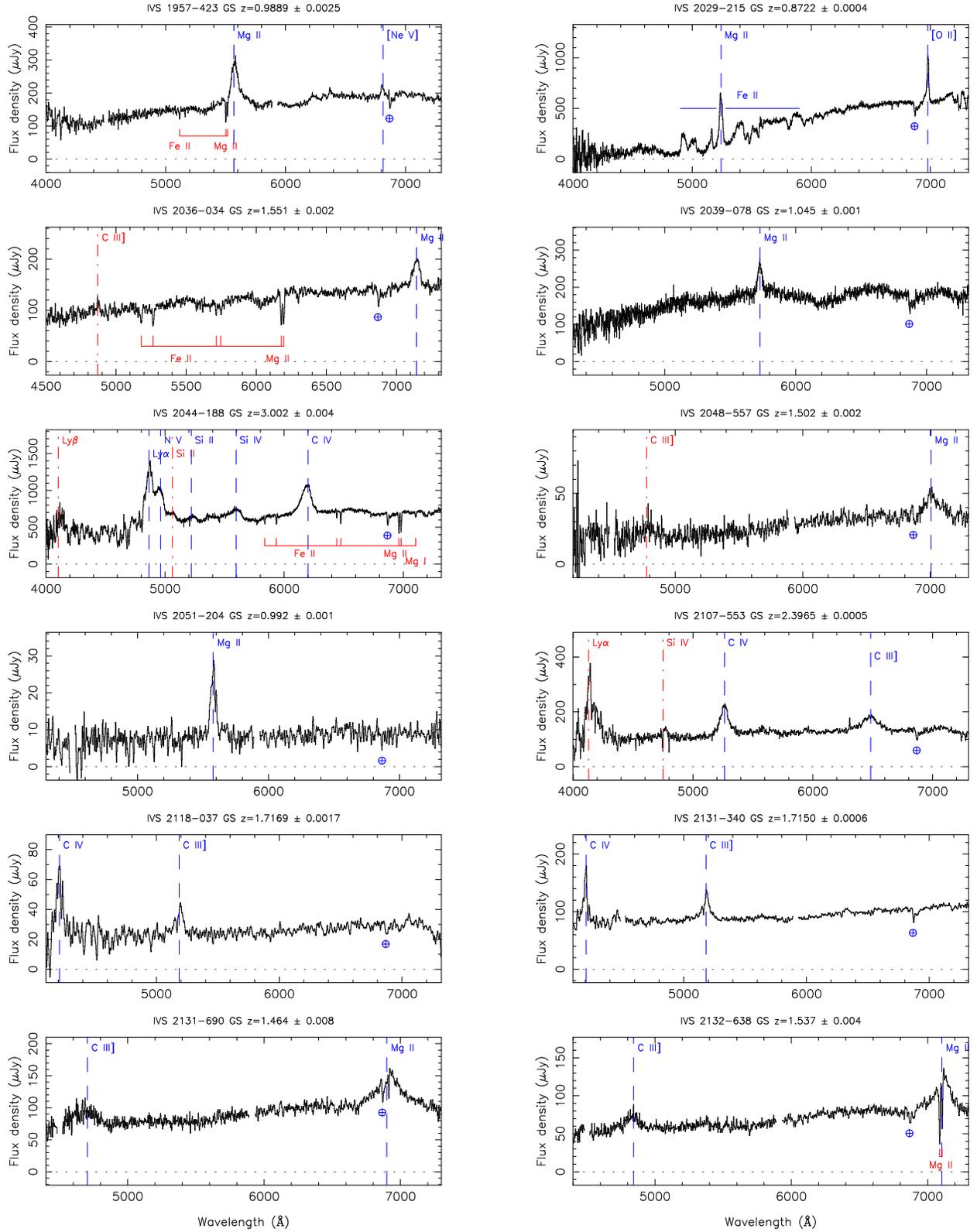

Fig. 1 (continued).—

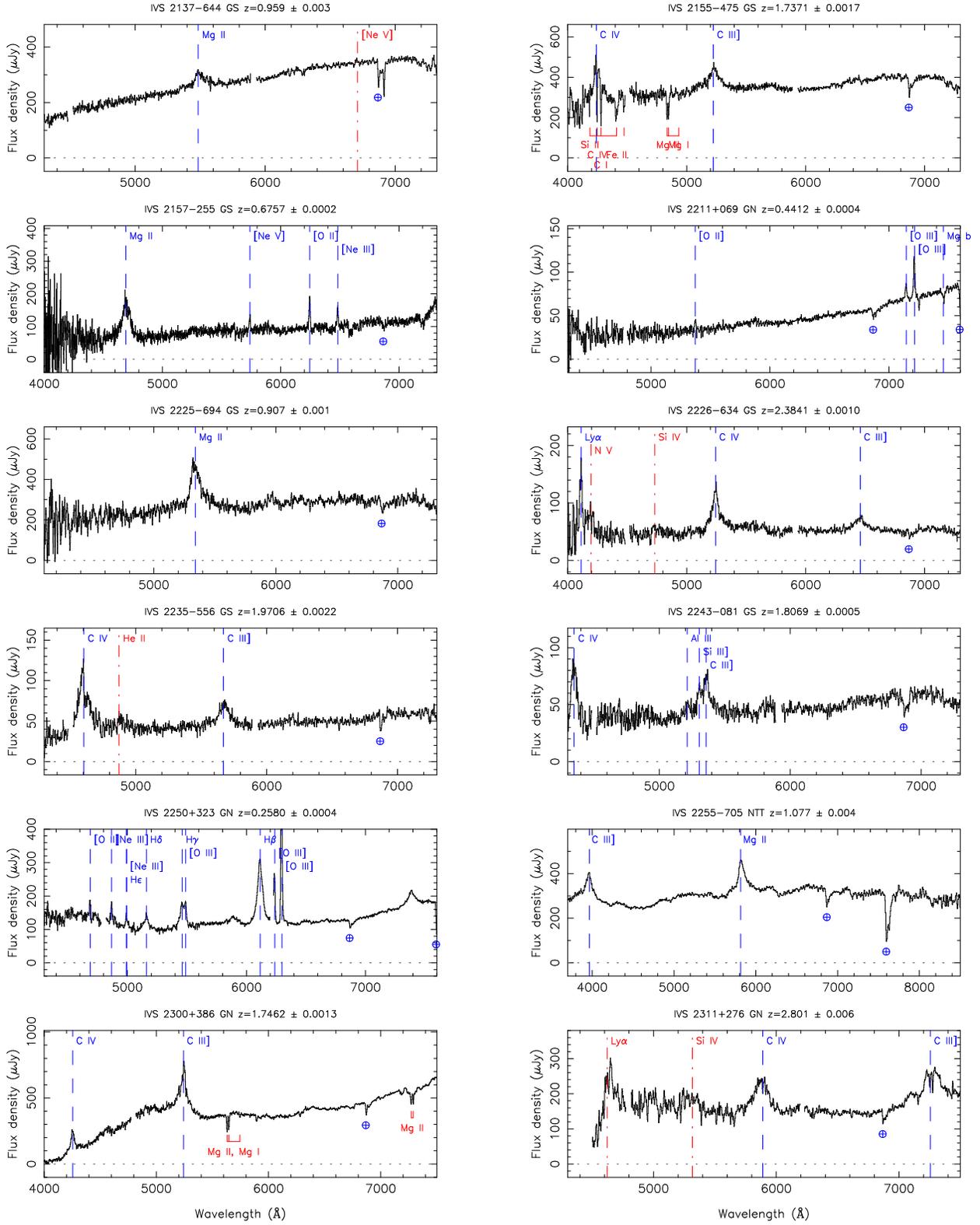

Fig. 1 (continued).—

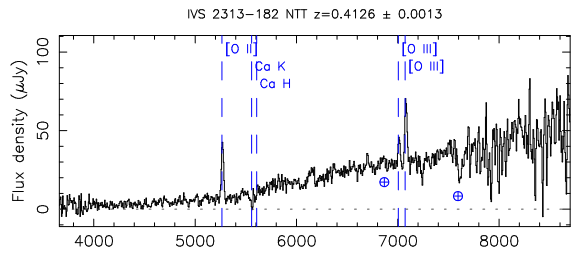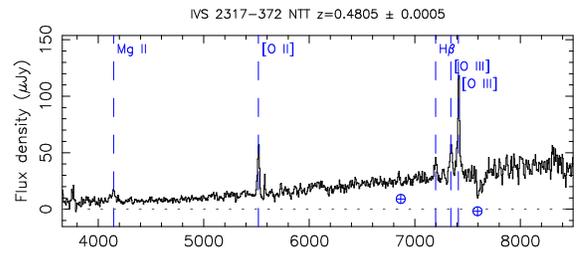

Fig. 1 (continued).—

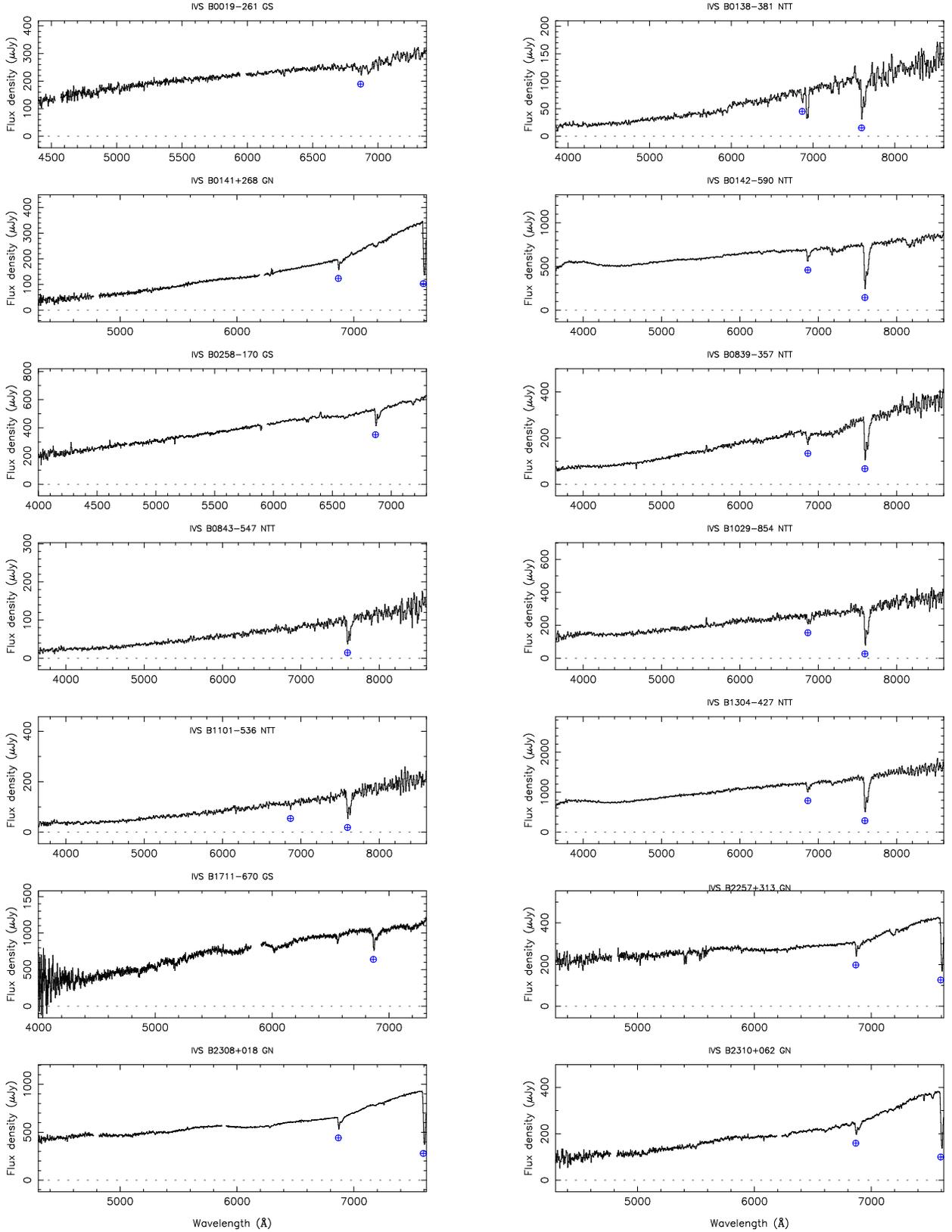

Fig. 2.— Spectra of 14 probable BL Lac objects from our observations, classified on the basis of their featureless spectra; see Table 2 for ICRF2 positions.

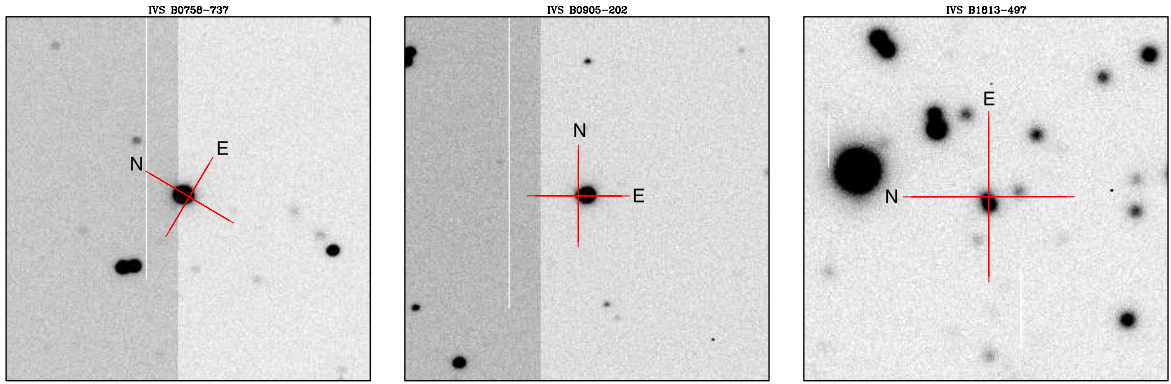

Fig. 3.— Acquisition images for the fields of IVS B0758−737, B0905−202, and B1813−497, all of which are partly obscured by a foreground star; position offsets between the radio source and obscuring star are 1.3″, 1.6″ and 0.8″, respectively. The arms of the cross marking the radio position are all 20 arcsec, and the orientation of each finding chart is indicated. The step in contrast in the two leftmost NTT images is the result of fast readout using two amplifiers.

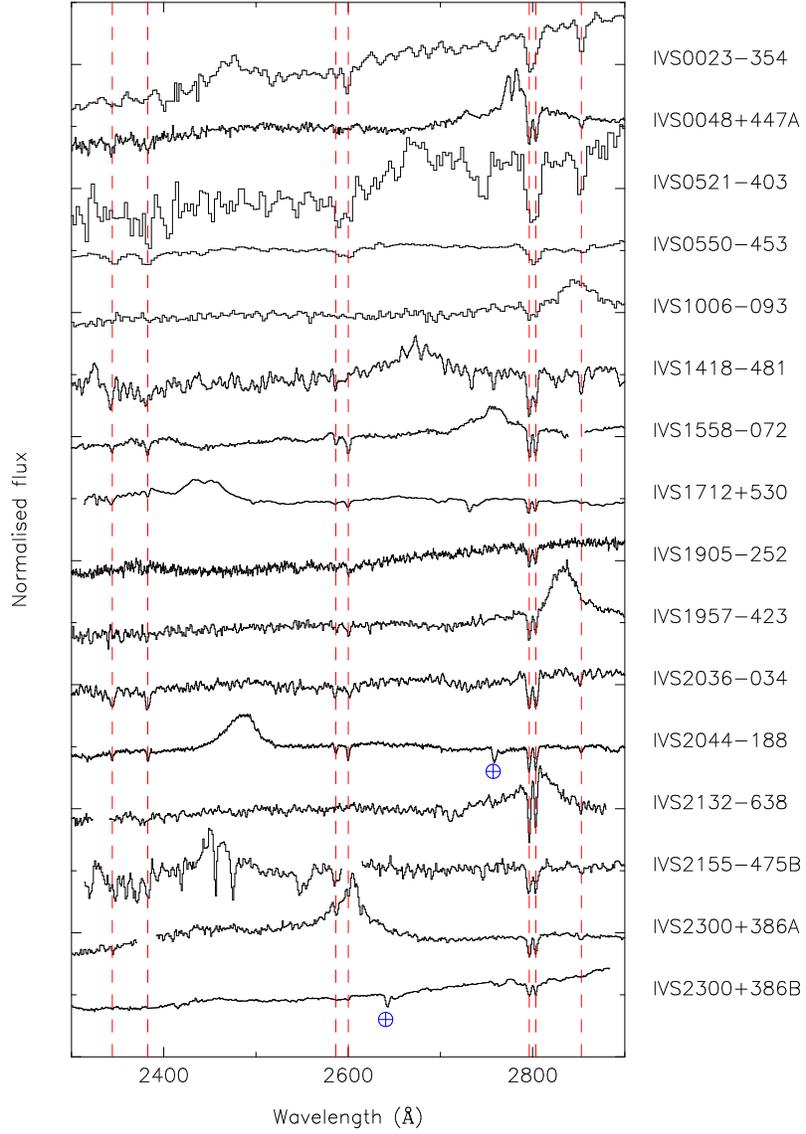

Fig. 4.— Montage of the Mg II absorption systems shifted back to the absorption rest frame. Telluric lines are marked and short spectral gaps correspond to gaps between the GMOS CCDs.

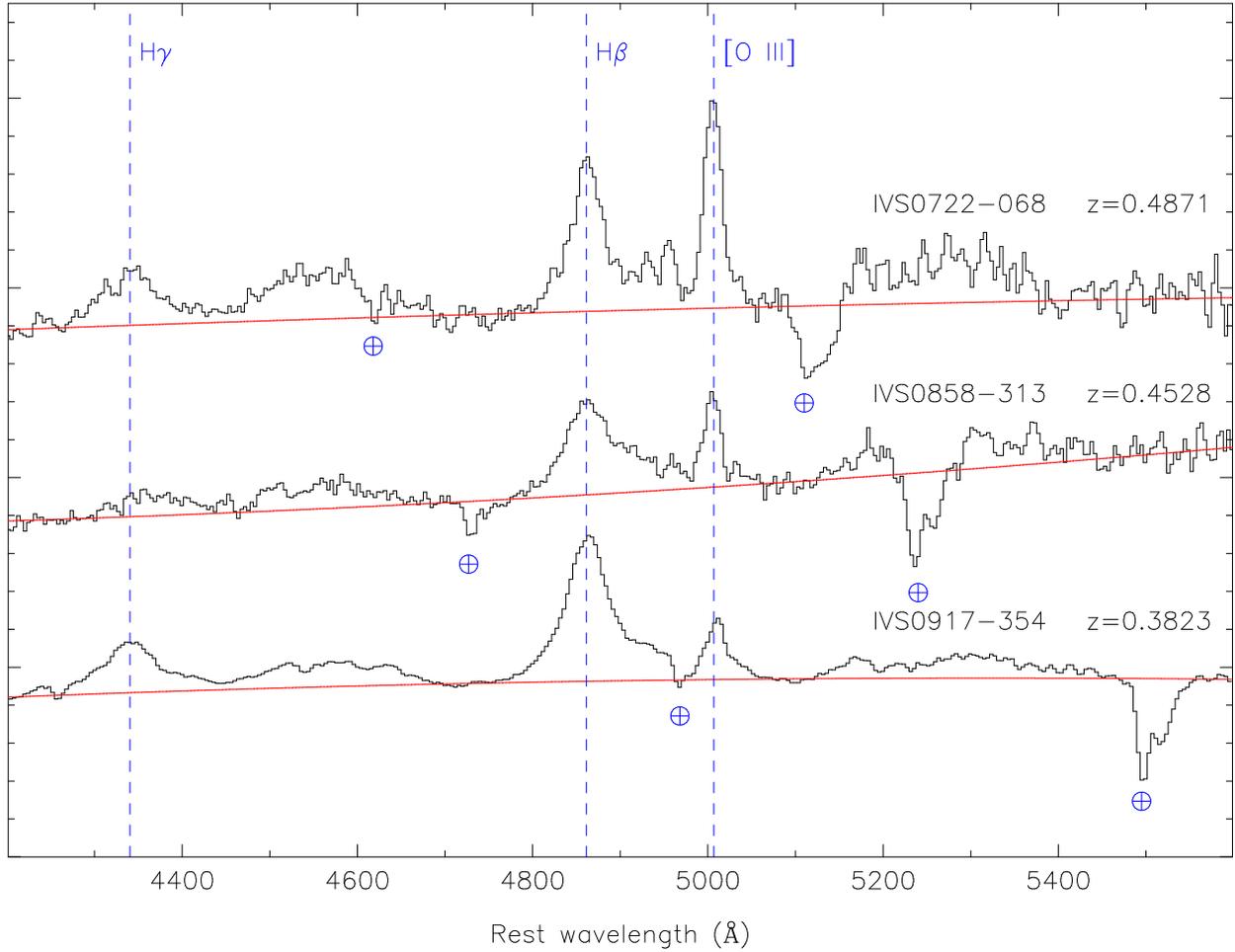

Fig. 5.— Montage of three $z \sim 0.4$ quasar spectra showing prominent broad Fe II emission both blueward and redward of Hβ. Spectra have been shifted to the rest frame and telluric features are marked. The effect of Fe II emission on the Hβ/[O III] profiles is most clearly seen in the high signal-to-noise spectrum of IVS B0917−354. The solid line (red) shows our estimated continuum level.

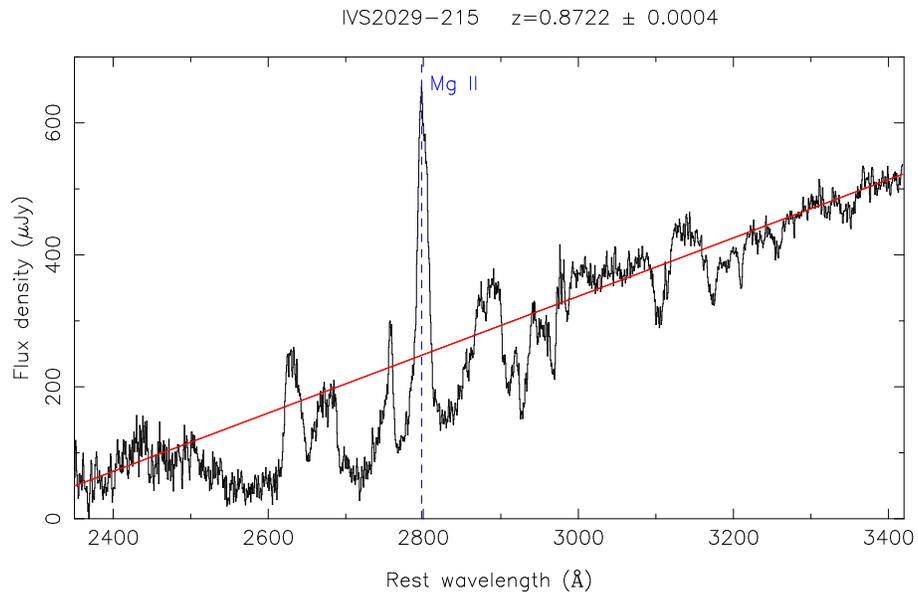

Fig. 6.— Optical spectrum of the $z = 0.8722$ FeLoBAL quasar IVS B2029−215 shifted to the emission rest frame. A tentative underlying continuum is shown (see text).